\documentclass[12pt,twoside]{article}

\usepackage{pslatex}	
\usepackage{amsfonts}
\usepackage{amsmath}
\usepackage{amsthm}
\usepackage{amscd}
\usepackage{cite}
\usepackage{epsf}
\usepackage{epsfig}
\usepackage{a4}

\def\beq{\begin{equation}}
\def\eeq{\end{equation}}
\def\bea{\begin{eqnarray}}
\def\eea{\end{eqnarray}}

\def\beann{\begin{eqnarray*}}
\def\eeann{\end{eqnarray*}}

\let\a=\alpha \let\be=\beta \let\g=\gamma \let\de=\delta
\let\e=\varepsilon \let\z=\zeta \let\h=\eta 

 \let\k=\kappa \let\la=\lambda \let\m=\mu
\let\n=\nu \let\x=\xi \let\p=\pi \let\r=\rho \let\s=\sigma
\let\om=\omega 
\let\ph=\varphi  \let\PH=\Phi 
  
\let\La=\Lambda  \let\D=\Delta

\let\qd=\quad \let\qqd=\qquad 

\def\epp{\, .}
\def\epc{\, ,}

\def\tst#1{{\textstyle #1}}
\def\dst#1{{\displaystyle #1}}

\theoremstyle{plain}
\newtheorem{theorem}{Theorem}
\newtheorem{lemma}{Lemma}

\newtheorem{corollary}{Corollary}
\newtheorem*{corollary*}{Corollary}

\theoremstyle{definition}

\def\2{\frac{1}{2}} \def\4{\frac{1}{4}}

\def\6{\partial}

\def\+{\dagger}

\def\<{\langle} \def\>{\rangle}

\let\then=\Rightarrow
\let\auf=\uparrow \let\ab=\downarrow

\def\i{{\rm i}}

\def\rd{{\rm d}}
\def\re{{\rm e}}

\DeclareMathOperator{\sh}{sh}
\DeclareMathOperator{\ch}{ch}

\DeclareMathOperator{\tr}{tr}

\DeclareMathOperator{\sign}{sign}
\DeclareMathOperator{\End}{End}

\DeclareMathOperator{\card}{card}

\def\fa{\mathfrak{a}}

\def\faq{\overline{\mathfrak{a}}}

\def\fb{\mathfrak{b}}

\def\aqq{\widetilde{\alpha}}
\def\bqq{\widetilde{\beta}}

\renewcommand{\appendix}{%
   \renewcommand{\section}{
	\secdef\Appendix\sAppendix}%
   \setcounter{section}{0}%
   \renewcommand{\thesection}{\Alph{section}}%
   \renewcommand{\theequation}{\thesection.\arabic{equation}}%
}

\newcommand{\Appendix}[2][?]{%
     \refstepcounter{section}%
     \setcounter{equation}{0}%
     \addcontentsline{toc}{appendix}%
          {\protect\numberline{\appendixname~\thesection} #1}%
     \vspace{\baselineskip}%
     {\noindent\large\bfseries\appendixname\ \thesection: #2\par}%
     \sectionmark{#1}\vspace{\baselineskip}}

\newcommand{\sAppendix}[1]{%
     {\noindent\large\bfseries\appendixname\:: #1\par}%
     \sectionmark{#1}\vspace{\baselineskip}}

\renewcommand{\tilde}{\widetilde}

\begin{document}

\thispagestyle{empty}

\begin{center}

{\Large {\bf Finite temperature density matrix and two-point
correlations in the antiferromagnetic XXZ chain\\}}

\vspace{7mm}

{\large Frank G\"{o}hmann\footnote[2]{e-mail:
goehmann@physik.uni-wuppertal.de},
Nils P. Hasenclever\footnote[1]{e-mail:
hasenclever@physik.uni-wuppertal.de}
and Alexander Seel%
\footnote[3]{e-mail: seel@physik.uni-wuppertal.de}\\

\vspace{5mm}

Fachbereich C -- Physik, Bergische Universit\"at Wuppertal,\\
42097 Wuppertal, Germany\\}

\vspace{20mm}

{\large {\bf Abstract}}

\end{center}

\begin{list}{}{\addtolength{\rightmargin}{9mm}
               \addtolength{\topsep}{-5mm}}
\item
We derive finite temperature versions of integral formulae for
the two-point correlation functions in the antiferromagnetic XXZ chain.
The derivation is based on the summation of density matrix elements
characterizing a finite chain segment of length $m$. On this occasion
we also supply a proof of the basic integral formula for the density
matrix presented in an earlier publication.
\\[2ex]
\end{list}

\clearpage

\section{Introduction}
Although many Yang-Baxter integrable quantum systems \cite{KBIBo,%
thebook} with important applications in diverse areas of physics
have been discovered and intensely studied since the 60s, nowadays
knowledge about the correlation functions of local observables in these
models still mostly relies on approximate methods like the application
of conformal field theories for critical models or direct numerical
studies of finite systems. Any progress in the exact calculation of
correlation functions is hindered by the complicated structure of the
many-body wave functions, depending on parameters, the Bethe roots,
which, in most cases, are not known explicitly. But also in those cases,
where explicit expressions for the Bethe roots are available and the
full finite temperature dynamical correlation functions could be
calculated, namely for the impenetrable Bose \cite{KBIBo} and Fermi
\cite{GIKP98,GIK98} gases, rather non-elementary mathematics is
involved, and even the asymptotic form of the correlation functions
is rather complicated.

So far the only Yang-Baxter integrable system with a non-trivial
distribution of Bethe roots in the ground state for which some 
more or less explicit exact results on the correlation functions could
be obtained is the anisotropic Heisenberg antiferromagnet with
Hamiltonian
\begin{equation} \label{xxzham}
     H_{XXZ} = J \sum_{j=1}^L \Bigl( \s_{j-1}^x \s_j^x
                  + \s_{j-1}^y \s_j^y + \D (\s_{j-1}^z \s_j^z - 1)
		     \Bigr) \epp
\end{equation}
It acts on the tensor product of $L$ spins $\2$. The local
operators $\s^\a$, $\a = x, y, z$, in (\ref{xxzham}) are the Pauli
matrices and $J > 0$ and $\D > - 1$ are two real parameters, called
exchange interaction and exchange anisotropy.

Important progress in the calculation of the correlation functions
of the XXZ chain was achieved in 1992, when Jimbo et al.\ \cite{JMMN92}
shifted the focus of attention from the two-point functions to the
density matrix of a segment of length $m$ of the infinite chain.
From a physical perspective the density matrix of a sub-system
seems a rather complicated object. Yet, it can be used to express the
expectation value of any operator acting on sites 1 to $m$ and, for
Yang-Baxter integrable systems, has an interesting mathematical
structure, when generalized to include inhomogeneities on columns 1 to
$m$ of the six-vertex model associated with the Hamiltonian
(\ref{xxzham}). As was shown later \cite{BJMST04a,BJMST04b}, the
inhomogeneous density matrix elements satisfy a set of functional
equations in the inhomogeneity parameters which was termed the reduced
q-Knizhnik-Zamolodchikov equation (rqKZ equation).

In \cite{JMMN92,JiMi96} (multiple) integral formulae for the
inhomogeneous density matrix elements were obtained. These played a
decisive role in the further developments of the following years.
Their generalization to the finite temperature case proposed in
\cite{GKS05} will be proven below and is the starting point for our
derivation of integral formulae for two-point functions at finite
temperature. Our reasoning will be based on the algebraic Bethe
ansatz solution of the XXZ chain and will make extensive use of the
Yang-Baxter algebra. Such type of approach was pioneered in
\cite{KMT99b}, where the integral formula for the density matrix
of \cite{JMMN92,JiMi96} was rederived and a longitudinal magnetic field,
not accessible by the original symmetry based approach of
\cite{JMMN92,JiMi96} was incorporated into the calculation.

At least at zero temperature the integral formulae for the density
matrix turned out to be a means to derive completely explicit results
for short-range correlation functions. This was first demonstrated in
the isotropic XXX case \cite{BoKo01} ($\D = 1$ in (\ref{xxzham})) with
the example of a special matrix element of the density matrix, called
the emptiness formation probability \cite{KIEU94}. In succeeding articles
\cite{BoKo02,BKNS02,SSNT03} further explicit examples of short range
correlation functions were obtained by calculating integrals, and the
technique of \cite{BoKo01} was extended to the anisotropic case
\cite{KSTS03,KSTS04,TKS04}. The knowledge of a number of explicit
examples then helped to develop an ansatz \cite{BKS03,BKS04a} for the
general inhomogeneous density matrix element in the XXX case which did
not involve integrals anymore. This was subsequently proved and further
generalized and simplified \cite{BJMST04a,BJMST04b,BJMST05a}. Finally
it was brought into a form which suits for taking the homogeneous limit
\cite{BJMST05b}.

The `Bethe ansatz approach' of \cite{KMT99b} was further developed in
\cite{KMST02a}, where integral formulae for two point functions and
for a certain one-parameter generating function of the $S^z$-$S^z$
correlation functions were obtained. The important feature of these
formulae is that they are different from those obtained by summing
the integrals for the density matrix elements naively, in that they
involve only a sum over as many (multiple) integrals as lattice sites,
while the number of terms in the naive sum is growing exponentially.
In a sense the formulae of \cite{KMST02a} give an implicit solution
of the problem of re-summing the naive sum into larger blocks. This
probably inspired the work \cite{KMST04c} on the explicit re-summation
of sums over density matrix elements. Note, however, that the formulae
of \cite{KMST04c} differ from those in \cite{KMST02a}. They seem to
be simpler and led to new explicit results for $\D = \2$ in
\cite{KMST05pp}.

This work is on the generalization of \cite{KMT99b} and \cite{KMST04c}
to finite temperatures. It is based on previous joint work of two of the
authors with A. Kl\"umper, where the quantum transfer matrix approach
\cite{Suzuki85,SuIn87,Kluemper92,Kluemper93} to the thermodynamics of
integrable lattice models was combined with certain combinatorial
results \cite{KBIBo,Slavnov89,KMST02a} from the algebraic Bethe ansatz.
As a result integral formulae for two-point functions \cite{GKS04a} and
density matrix elements \cite{GKS05,GKS05b} at finite temperature and
finite longitudinal magnetic field were obtained. In this work we shall
present a proof of the integral formula \cite{GKS05} in the general case.
We further apply the summation formulae of \cite{KMST04c} to the
finite temperature density matrix and obtain new integral
representations for two-point functions at finite temperatures.

In the next section we review the algebraic Bethe ansatz approach to
finite temperature correlation functions \cite{GKS04a}. Section 3 is
devoted to the density matrix. We explain its origin in physics and show
how it naturally fits into the formalism of section 2. The remaining
part of the section contains a proof of the multiple integral formula
for the density matrix elements proposed in \cite{GKS05} and shown in
theorem \ref{theo:base}. Some technical parts of the proof are deferred
to appendix \ref{app:leftact}. In section 4 and appendix B we show how to
sum up the density matrix elements into efficient formulae for the
generating function of the $S^z$-$S^z$ correlation function and for the
correlation function $\< \s_1^+ \s_{m+1}^-\>_{T, h}$. We conclude with
a summary and an outlook in section 5.
\enlargethispage{\baselineskip}

\section{Algebraic Bethe ansatz approach to finite temperature
correlation functions}
Before coming to the definition of the density matrix and to its
representations as expectation value of non-local operators in a
certain Bethe ansatz state and as a multiple integral we have to
fix some basic notation. We follow our earlier paper \cite{GKS04a},
where more details and proofs of most of the statements cited below
can be found.

In the framework of Yang-Baxter integrability the XXZ chain with
Hamiltonian (\ref{xxzham}) is the fundamental model \cite{thebook}
associated with the solution $R(\la,\m) \in \End \bigl( {\mathbb C}^2
\otimes {\mathbb C}^2 \bigr)$ of the Yang-Baxter equation
\begin{equation} \label{ybe}
     R_{\a' \be'}^{\a \be} (\la, \m) R_{\a'' \g'}^{\a' \g} (\la, \n)
     R_{\be'' \g''}^{\be' \g'} (\m, \n) =
     R_{\be' \g'}^{\be \g} (\m, \n) R_{\a' \g''}^{\a \g'} (\la, \n)
     R_{\a'' \be''}^{\a' \be'} (\la, \m)
\end{equation}
defined by
\begin{align} \label{rxxz}
     R(\la,\m) & = \begin{pmatrix}
                    1 & 0 & 0 & 0 \\
		    0 & b(\la, \m) & c(\la, \m) & 0 \\
		    0 & c(\la, \m) & b(\la, \m) & 0 \\
		    0 & 0 & 0 & 1
		   \end{pmatrix} \epc \\[2ex] \label{defbc}
     b(\la, \m) & = \frac{\sh(\la - \m)}{\sh(\la - \m + \h)} \epc \qd
     c(\la, \m) = \frac{\sh(\h)}{\sh(\la - \m + \h)} \epp
\end{align}

For the exact calculation of thermal averages at temperature $T$ we
need to express the statistical operator of the canonical ensemble
\begin{equation} \label{statop}
      \r_L = \re^{- H_{XXZ}/T}
\end{equation}
in terms of the $R$-matrix (\ref{rxxz}) in a way that allows us to
utilize the Yang-Baxter algebra (see (\ref{yba}) below) efficiently.
For this purpose we introduce an auxiliary vertex model defined by a
monodromy matrix
\begin{equation} \label{monoqtm}
     T_j (\la ) = \begin{pmatrix} A(\la) & B(\la) \\ C(\la) & D(\la)
                  \end{pmatrix}_j =
        R_{j \overline{N}} \bigl(\la, \tst{\frac{\be}{N}} \bigr)
	R_{\overline{N-1} j}^{t_1}
	   \bigl(- \tst{\frac{\be}{N}}, \la \bigr) \dots
        R_{j \overline{2}} \bigl(\la, \tst{\frac{\be}{N}} \bigr)
	R_{\bar 1 j}^{t_1} \bigl(- \tst{\frac{\be}{N}}, \la \bigr) \epp
\end{equation}
This monodromy matrix acts non-trivially in the tensor product of an
auxiliary space $j = 1, \dots, L$ and $N \in 2 {\mathbb N}$ `quantum
spaces' $\bar 1, \dots, \overline{N}$ which are all isomorphic to
${\mathbb C}^2$. By $t_1$ we mean transposition with respect to the
first of the two spaces into which $R(\la,\m)$ is embedded. The
parameter $\be$ is inversely proportional to the temperature $T$. By
definition,
\begin{equation} \label{defbeta}
     \be = \frac{2J \sh(\h)}{T} \epp
\end{equation}

The monodromy matrix (\ref{monoqtm}) has many remarkable properties.
First of all it can be used to express the statistical operator
(\ref{statop}) as a limit,
\begin{equation} \label{staatsoper}
     \r_L = \lim_{N \rightarrow \infty}
            \tr_{\bar 1 \dots \bar N} T_1 (0) \dots T_L (0) \epc
\end{equation}
with the implicit identification $\D = \ch(\h)$. The number $N$ in
(\ref{staatsoper}) is called the Trotter number, and the limit is
referred to as the Trotter limit.

For large Trotter numbers the operator
\begin{equation}
     \r_{N, L} = \tr_{\bar 1 \dots \bar N} T_1 (0) \dots T_L (0)
\end{equation}
is a good approximation to the statistical operator. The thermal
expectation value of a product of local operators $X^{(n)}$ acting on
sites $j, \dots, j + m - 1$ of the spin chain is then approximated by
\begin{multline} \label{finitecor}
     \< X_j^{(1)} \dots X_{j + m - 1}^{(m)} \>_{N, T}
        = \frac{\tr_{1 \dots L} \r_{N, L} X_j^{(1)} \dots
	                                  X_{j + m - 1}^{(m)}}
               {\tr_{1 \dots L} \r_{N, L}} \\
        = \frac{\tr_{\bar 1 \dots \bar N}
	        \bigl( \tr T(0) \bigr)^{j-1}
	        \tr X^{(1)} T(0) \dots \tr X^{(m)} T(0)
	        \bigl( \tr T(0) \bigr)^{L-m-j+1}}
               {\tr_{\bar 1 \dots \bar N} \bigl( \tr T(0) \bigr)^L} \epp
\end{multline}
Here the expression
\begin{equation} \label{finitepart}
     Z_{N, L} = \tr_{1 \dots L} \r_{N, L}
              = \tr_{\bar 1 \dots \bar N} \bigl( \tr T(0) \bigr)^L
\end{equation}
in the denominator is the finite Trotter number approximant to the
partition function $Z_L = \tr_{1 \dots L} \exp (- H_{xxz}/T)$ of the
XXZ chain of length $L$.

The operator occurring on the right hand side of equation
(\ref{finitepart}),
\begin{equation} \label{tqtm}
     t(\la) = \tr T (\la) = A(\la) + D(\la) \epc
\end{equation}
is called the quantum transfer matrix. By construction it can be
diagonalized by means of the algebraic Bethe ansatz. This follows
\cite{KBIBo} since the monodromy matrix (\ref{monoqtm}) defines a
representation of the Yang-Baxter algebra,
\begin{equation} \label{yba}
    R_{jk} (\la,\m) T_j (\la) T_k (\m)
       = T_k (\m) T_j (\la) R_{jk} (\la,\m) \epc
\end{equation}
and has a pseudo vacuum $|0\> = \bigl[\bigl( \begin{smallmatrix} 0 \\ 1
\end{smallmatrix} \bigr) \otimes \bigl( \begin{smallmatrix} 1 \\ 0
\end{smallmatrix} \bigr)\bigr]^{\otimes N/2}$ on which it acts like
\begin{equation}
\begin{split}
     C(\la) |0\> & = 0 \epc \\
     A(\la) |0\> & = b^\frac{N}{2}
                     \Bigr(- \tst{\frac{\be}{N}}, \la \Bigr)
		     |0\> = a(\la) |0\> \epc \qd 
     D(\la) |0\> = b^\frac{N}{2} \Bigr(\la, \tst{\frac{\be}{N}} \Bigr)
                   |0\> = d(\la) |0\> \epp
\end{split}
\end{equation}
Then (see \cite{KBIBo}) the vector
\begin{equation} \label{abaevec}
     |\{\la\}\> = |\{\la_j\}_{j=1}^M\> = B(\la_1) \dots B(\la_M) |0\>
\end{equation}
is an eigenvector of the quantum transfer matrix $t (\la)$ if
the Bethe roots $\la_j$, $j = 1, \dots, M$, satisfy the system
\begin{equation} \label{bae}
     \frac{a(\la_j)}{d(\la_j)}
        = \prod_{\substack{k = 1 \\k \ne j}}^M
	  \frac{\sh(\la_j - \la_k + \h)}{\sh(\la_j - \la_k - \h)}
\end{equation}
of Bethe ansatz equations. The corresponding eigenvalue is
\begin{equation} \label{abaeval}
     \La(\la) = a(\la) \prod_{j=1}^M \frac{\sh(\la - \la_j - \h)}
                                          {\sh(\la - \la_j)}
	      + d(\la) \prod_{j=1}^M \frac{\sh(\la - \la_j + \h)}
		                          {\sh(\la - \la_j)} \epp
\end{equation}
Note that the `creation operators' $B(\la_j)$ mutually commute.
Therefore the eigenvector (\ref{abaevec}) is symmetric in the
Bethe roots, whence our notation $|\{\la\}\>$.

Equations (\ref{bae}), (\ref{abaeval}) are a means to study the spectrum
of the quantum transfer matrix (\ref{tqtm}). It is easy to see that for
every finite Trotter number $N$ and high enough temperature $T$ the
quantum transfer matrix has a non-degenerate eigenvalue with largest
modulus at $\la = 0$ which is characterized by a unique set $\{\la\}
= \{\la_j\}_{j=1}^{N/2}$ of $M = N/2$ Bethe roots that all go to zero
as $T$ goes to infinity. We call it the dominant eigenvalue and denote
it by $\La_0 (\la)$. The corresponding eigenvector $|\{\la\}\>$ will
be called the dominant eigenvector. It is a crucial property of the
whole construction that the dominant eigenvalue alone determines the
thermodynamic properties of the XXZ chain in the thermodynamic limit
$L \rightarrow \infty$. This can be seen as follows. The free energy per
lattice site in the thermodynamic limit follows from (\ref{finitepart})
as
\begin{equation} \label{freedoublelimit}
     f(T) = - T \lim_{L \rightarrow \infty} \frac{1}{L}
                \lim_{N \rightarrow \infty}
		\ln \tr_{\bar 1 \dots \bar N} t^L (0) \epp
\end{equation}
But the two limits in (\ref{freedoublelimit}) commute according to
\cite{Suzuki85,SuIn87}, hence
\begin{equation} \label{freesinglelimit}
     f(T) = - T \lim_{N \rightarrow \infty} \ln \La_0 (0) \epp
\end{equation}
Applying a similar reasoning to the expectation value of the product
of local operators (\ref{finitecor}) in the thermodynamic limit we obtain
\cite{GKS04a}
\begin{equation} \label{cor1}
     \< X_j^{(1)} \dots X_{j + m - 1}^{(m)} \>_{T}
        = \lim_{N \rightarrow \infty}
	  \frac{\<\{\la\}| \tr X^{(1)} T(0) \dots \tr X^{(m)} T(0)
	        |\{\la\}\>}{\<\{\la\}|\{\la\}\> \La_0^m (0)} \epc
\end{equation}
where $|\{\la\}\>$ denotes the dominant eigenvector. This means that
the dominant eigenvector determines all thermal correlation functions
of the system in the thermodynamic limit. In other words, it determines
the state of thermal equilibrium completely.

The expression under the limit on the right hand side of (\ref{cor1})
expresses the finite Trotter number approximant of the expectation
value of the product of operators on the left hand side of the equation
as an expectation value of monodromy matrix elements in a Bethe state.
But this is still not good enough for applying the Yang-Baxter algebra
(\ref{yba}) since the arguments of the monodromy matrix elements are
all zero, and the relations in the Yang-Baxter algebra for commuting two
monodromy matrix elements with the same spectral parameter $\la$ are
singular. Therefore we first regularize the expectation value by
introducing a set of mutually distinct `inhomogeneities' $\x_1, \dots,
\x_m$ and then calculate the correlation functions by means of the
formula
\begin{equation} \label{cor2}
     \< X_j^{(1)} \dots X_{j + m - 1}^{(m)} \>_{T}
        = \lim_{N \rightarrow \infty}
	  \lim_{\substack{\x_j \rightarrow 0\\j = 1, \dots, m}}
	  \frac{\<\{\la\}| \tr X^{(1)} T(\x_1) \dots \tr X^{(m)} T(\x_m)
	                   |\{\la\}\>}
	       {\<\{\la\}|\{\la\}\> \prod_{j=1}^m \La_0 (\x_j)} \epp
\end{equation}

Since the Hamiltonian (\ref{xxzham}) commutes with the $z$-component
\begin{equation}
     S^z = \tst{\2} \sum_{j = 1}^L \s_j^z
\end{equation}
of the total spin, thermodynamics and finite temperature correlation
functions can still be treated within the `quantum transfer matrix
approach' described above if the system is exposed to an external
magnetic field $h$ in $z$-direction \cite{GKS04a}. The external field
is properly taken into account by applying a twist to the monodromy
matrix (\ref{monoqtm}),
\begin{equation}
     T (\la) \rightarrow
        T (\la) \Bigl( \begin{smallmatrix} \re^{h/2T} & 0 \\
	               0 & \re^{- h/2T} \end{smallmatrix} \Bigr) \epp
\end{equation}

In \cite{Kluemper92,Kluemper93} techniques were developed to treat the
Trotter limit in (\ref{freesinglelimit}) analytically. The main idea
is to associate with every Trotter number $N \in 2 {\mathbb N}$ a
meromorphic `auxiliary function' 
\begin{equation} \label{defa}
     \fa_N (\la) = \frac{d(\la)}{a(\la)} \prod_{k = 1}^{N/2}
	           \frac{\sh(\la - \la_k + \h)}{\sh(\la - \la_k - \h)}
		   \epc
\end{equation}
where the $\la_j$ are the Bethe roots that characterize the dominant
eigenvalue, and to consider the sequence of these functions in the
complex plane, rather than sequences of sets of Bethe roots. This is
recommendable, since the distribution of Bethe roots for large Trotter
number $N$ is not smooth but develops a limit point at $\la = 0$.
The sequence of functions $\fa_N$, on the other hand, converges to a
limit function $\fa$ which is analytic far enough from $\la = 0$. The
strategy is thus to express all physical quantities of interest as
contour integrals over $\fa_N$ on contours far away from the origin
and then take the Trotter limit.
\enlargethispage{\baselineskip}

In particular $\ln \fa_N$ can be expressed this way \cite{GKS04a},
which yields a non-linear integral equation for $\fa_N$,
\begin{multline} \label{nlien}
     \ln \fa_N (\la) = - \frac{h}{T} + 
                      \ln \biggl[
                      \frac{\sh(\la - \frac{\be}{N})
		            \sh(\la + \frac{\be}{N} + \h)}
			   {\sh(\la + \frac{\be}{N})
			    \sh(\la - \frac{\be}{N} + \h)}
			    \biggr]^{\mspace{-3mu} \frac{N}{2}} \\
        - \int_{\cal C} \frac{d \om}{2 \p \i} \,
	  \frac{\sh (2 \h) \ln (1 + \fa_N (\om))}
	       {\sh(\la - \om + \h) \sh(\la - \om - \h)} \epp
\end{multline}
Notice that we included the
\begin{figure}

\begin{center}

\epsfxsize \textwidth
\epsffile{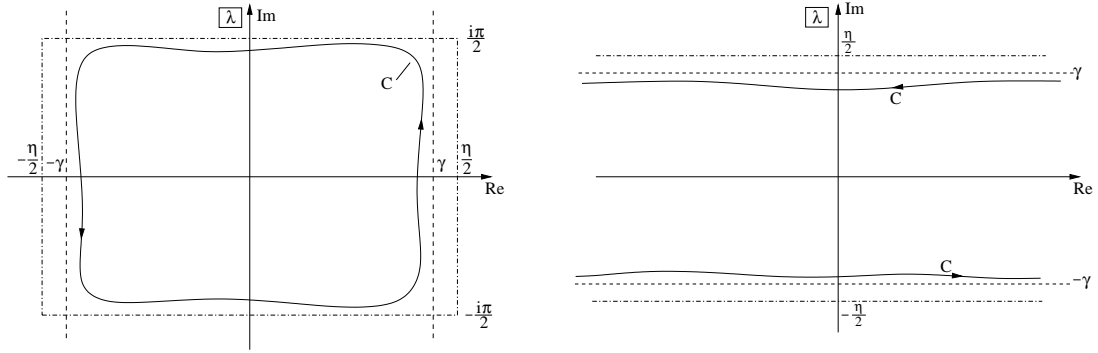}

\caption{\label{fig:cancon} The canonical contour ${\cal C}$ in the
off-critical regime $\D > 1$ (left panel) and in the critical regime
$|\D | < 1$ (right panel). For $\D > 1$ the contour is a rectangle
with sides at $\pm \i \frac{\p}{2}$ and $\pm \g$, where $\g$ is
slightly smaller than $\frac{\h}{2}$. For $|\D| < 1$ the contour
surrounds the real axis at a distance $|\g|$ slightly less than
$\frac{|\h|}{2}$.}
\end{center}
\end{figure}
magnetic field here according to the recipe explained above.
The integration contour ${\cal C}$ depends on $\h$ (see
figure~\ref{fig:cancon}). As is very important, however, it does not
depend on $N$. The Trotter number $N$ merely appears as a parameter in
the inhomogeneity of the integral equation. Hence, the Trotter limit
can easily be performed. Using also (\ref{defbeta}) we obtain
\begin{equation} \label{nlie}
     \ln \fa (\la) = - \frac{h}{T}
                     - \frac{2J \sh^2 (\h)}{T \sh(\la) \sh(\la + \h)}
                     - \int_{\cal C} \frac{d \om}{2 \p \i} \,
	               \frac{\sh (2 \h) \ln (1 + \fa (\om))}
		            {\sh(\la - \om + \h)
			            \sh(\la - \om - \h)} \epp
\end{equation}

A physical interpretation of the auxiliary function $\fa$ stems from
the `free Fermion limit' $\D \rightarrow 0$ (see e.g.\ \cite{GoSe05pp}),
where the Hamiltonian (\ref{xxzham}) describes the highly degenerate
XX chain. In this limit the function $1/(1 + \fa)$ becomes the Fermi
function. Thus, we may think of $1/(1 + \fa)$ as generalizing the
Fermi function to the interacting case. Thinking in terms of Fermions
it is natural to set $\faq = 1/\fa$, and interpret $1/(1 + \faq)$
as the Fermi function for holes satisfying $1/(1 + \fa) + 1/(1 + \faq)
= 1$.

In the thermodynamic limit the free energy per lattice site as a
function of temperature and magnetic field can be expressed as an
integral, involving either the auxiliary function $\fa$ or its
reciprocal $\faq$, over the same contour as in (\ref{nlie})
\cite{Kluemper04,GKS04a}
\begin{equation} \label{freee}
     \mspace{-3.mu}
     f (h,T) = - \frac{h}{2} - T \mspace{-2.mu}
                 \int_{\cal C} \frac{d \om}{2 \p \i} \,
                 \frac{\sh (\h) \ln (1 + \fa (\om))}
	              {\sh(\om) \sh(\om + \h)}
             = \frac{h}{2} + T \mspace{-2.mu}
	       \int_{\cal C} \frac{d \om}{2 \p \i} \,
               \frac{\sh (\h) \ln (1 + \faq (\om))}
	            {\sh(\om) \sh(\om - \h)}.
\end{equation}

\section{The density matrix} \label{sec:densmat}
The density matrix is a means to describe a sub-system as part of a
larger system in thermodynamic equilibrium in terms of the degrees of
freedom of the sub-system. Roughly speaking it is obtained by taking
the (normalized) statistical operator and tracing out the `unwanted'
degrees of freedom. It can be defined rather generally, but we content
ourselves here with the example when the sub-system consists of the
first $m$ sites of the XXZ chain of length $L > m$. Since we want to
include a magnetic field in $z$-direction into our considerations we
have to modify the statistical operator (\ref{statop}) as
\begin{equation} \label{statoph}
      \r_L = \exp \bigl( - (H_{XXZ} - h S^z)/T \bigr) \epp
\end{equation}
Then the density matrix of the sub-system consisting of the first
$m$ lattice sites is defined as
\begin{equation} \label{defdensmatgen}
     D_L (T, h) = \frac{\tr_{m+1 \dots L} \r_L}
                       {\tr_{1 \dots L} \r_L} \epp
\end{equation}
By construction, the thermal average of every operator $A$ acting
non-trivially only on sites 1 to $m$ can now be written as
\begin{equation}
     \<A\>_{T, h} = \frac{\tr_{1 \dots L} A \r_L}{\tr_{1 \dots L} \r_L}
                  = \frac{\tr_{1 \dots m} A_{1 \dots m}
		          \tr_{m+1 \dots L} \r_L}
		         {\tr_{1 \dots L} \r_L}
                  = \tr_{1 \dots m} A_{1 \dots m} D_L (T, h) \epc
\end{equation}
where $A_{1 \dots m}$ is the restriction of $A$ to a chain consisting
of sites 1 to $m$. In particular, every two-point function of local
operators in the segment 1 to $m$ of the XXZ chain can be brought into
the above form.

In order to make contact with the formalism explained in the previous
section we shall refer to a particular local basis. For our purpose it
is most convenient to choose the gl(2) standard basis consisting of the
$2 \times 2$ matrices $e^\a_\be$, $\a, \be = 1, 2$, with a single
non-zero entry at the intersection of row $\be$ and column $\a$. The
canonical embedding of these matrices into the space of operators on
the space of states $({\mathbb C}^2)^{\otimes L}$ of the Hamiltonian
(\ref{xxzham}) will be denoted by ${e_j}^\a_\be$, $j = 1, \dots, L$.
For their restriction to the first $m$  sites we use the same symbols.
Then we find the expressions
\begin{equation} \label{densmatell}
     {D_L}^{\a_1 \dots \a_m}_{\be_1 \dots \be_m} (T, h)
        = \tr_{1 \dots m} {e_1}^{\a_1}_{\be_1} \dots
                          {e_m}^{\a_m}_{\be_m} D_L (T, h)
        = \<{e_1}^{\a_1}_{\be_1} \dots {e_m}^{\a_m}_{\be_m}\>_{T, h}
\end{equation}
for the matrix elements of the density matrix, which hold for all $L$
and are still valid in the thermodynamic limit. To denote the
density matrix in the thermodynamic limit we simply leave out the
subscript $L$ in (\ref{densmatell}).

For $L \rightarrow \infty$ the right hand side of (\ref{densmatell})
is of the form of the left hand side of (\ref{cor2}) with $j = 1$ and
$X^{(n)} = e^{\a_n}_{\be_n}$. Hence, it can be calculated as
\begin{equation} \label{defdensmat}
     \bigl\<{e_1}_{\be_1}^{\a_1} \dots
            {e_m}_{\be_m}^{\a_m}\bigr\>_{T, h} =
        \lim_{N \rightarrow \infty} \:
	\lim_{\x_1, \dots, \x_m \rightarrow 0}
        {D^{(N)}}^{\a_1 \dots \a_m}_{\be_1 \dots \be_m}
	         (\x_1, \dots, \x_m) \epc
\end{equation}
where
\begin{equation} \label{defdn}
     {D^{(N)}}^{\a_1 \dots \a_m}_{\be_1 \dots \be_m} (\x_1, \dots, \x_m)
        = \frac{\< \{\la\}| T^{\a_1}_{\be_1} (\x_1)
	                     \dots T^{\a_m}_{\be_m} (\x_m)|\{\la\}\>}
             {\<\{\la\}| \prod_{j=1}^m t (\x_j) |\{\la\}\>}
\end{equation}
is the `inhomogeneous finite Trotter number approximant' to the density
matrix element (\ref{densmatell}) in the thermodynamic limit. In order
to simplify the notation we have suppressed the dependence on $T$ and
$h$ on the left hand side. Notice the formal similarity between
(\ref{defdensmatgen}) and (\ref{defdn}). In both cases the denominator
is the trace of the term in the numerator.

In \cite{GKS05} a basic integral formula for the inhomogeneous finite
Trotter number approximant (\ref{defdn}) of the density matrix elements
was proposed. The formula involves certain positive integers $\aqq_j^+$
and $\bqq_j^-$ derived from the sequences of indices $(\a_n)_{n=1}^m$,
$(\be_n)_{n=1}^m$ characterizing the matrix element. The indices take
values $1, 2$ corresponding to spin-up or spin-down. We shall denote
the position of the $j$th up-spin in $(\a_n)_{n=1}^m$ by $\a_j^+$ and
that of the $k$th down-spin in $(\be_n)_{n=1}^m$ by $\be_k^-$. Then,
by definition, $\aqq_j^+ = \a_{|\a^+| - j + 1}^+$, where $|\a^+|$ is
the number of up-spins in $(\a_n)_{n=1}^m$, and $j = 1, \dots, |\a^+|$.
Similarly $\bqq_j^- = \be_{k - |\a^+|}^-$, $k = |\a^+| + 1, \dots,
|\a^+| + |\be^-|$, with $|\be^-|$ denoting the number of down-spins
in $(\be_n)_{n=1}^m$.
\begin{theorem} \label{theo:base}
Due to the conservation of $S^z$ the matrix elements
${D^{(N)}}^{\a_1 \dots \a_m}_{\be_1 \dots \be_m} (\x_1, \dots, \x_m)$
vanish if $|\a^+| + |\be^-| \ne m$. For $|\a^+| + |\be^-| = m$ they are
given by the formula
\begin{align} \label{densint}
     {D^{(N)}}^{\a_1 \dots \a_m}_{\be_1 \dots \be_m}
               (\x_1, \dots, & \x_m)
        = \notag \\
	  \biggl[ \prod_{j=1}^{|\a^+|}
	     \int_{\cal C} & \frac{d \om_j}{2 \p \i (1 + \fa_N (\om_j))}
	     \prod_{k=1}^{\aqq_j^+ - 1} \sh(\om_j - \x_k - \h)
	     \prod_{k = \aqq_j^+ + 1}^m \sh(\om_j - \x_k) \biggr]
	     \notag \\
          \biggl[ \prod_{j = |\a^+| + 1}^{m}
	     \int_{\cal C} & \frac{d \om_j}{2 \p \i (1 + \faq_N (\om_j))}
	     \prod_{k=1}^{\bqq_j^- - 1} \sh(\om_j - \x_k + \h)
	     \prod_{k = \bqq_j^- + 1}^m \sh(\om_j - \x_k) \biggr]
	     \notag \\ &
        \frac{\det( - G(\om_j, \x_k))}
	     {\prod_{1 \le j < k \le m}
	         \sh(\x_k - \x_j) \sh( \om_j - \om_k - \h)} \epp
\end{align}
The function $G(\om,\x)$ depends on the Trotter number through $\fa_N$.
It has to be calculated from the linear integral equation
\begin{multline} \label{defg}
     G(\la,\x) = \frac{\sh(\h)}{\sh(\x - \la) \sh(\x - \la + \h)} \\
               + \int_{\cal C} \frac{d \om}{2 \p \i (1 + \fa_N (\om))} \,
	         \frac{\sh (2 \h)G (\om,\x)}
		      {\sh(\la - \om + \h)\sh(\la - \om - \h)} \epp
\end{multline}
The contour ${\cal C}$ in (\ref{densint}), (\ref{defg}) is the same as
in figure \ref{fig:cancon}.
\end{theorem}
Note that the function $G(\om,\x)$ introduced in \cite{GKS04a}
generalizes the `density function' $\r (\la)$, that determines the
ground state energy and the zero temperature magnetization, to finite
temperatures and to the inhomogeneous case. Taking the Trotter limit
and the homogeneous limit (for the latter compare \cite{KMT99b}) in
(\ref{densint}) we obtain an integral formula for the density matrix
elements (\ref{defdensmat}).
\begin{corollary}
Integral formula for the density matrix of a segment of length $m$ of
the XXZ chain in the thermodynamic limit.
\begin{align} \label{densinthom}
     \bigl\<{e_1}_{\be_1}^{\a_1} \dots
            {e_m}_{\be_m}^{\a_m}\bigr\>_{T, h} =
        \biggl[ \prod_{j=1}^{|\a^+|}
	     \int_{\cal C} & \frac{d \om_j}{2 \p \i (1 + \fa (\om_j))}
	     \sh^{\aqq_j^+ - 1} (\om_j - \h)
	     \sh^{m - \aqq_j^+} (\om_j) \biggr] \notag \\
        \biggl[ \prod_{j = |\a^+| + 1}^{m}
	     \int_{\cal C} & \frac{d \om_j}{2 \p \i (1 + \faq (\om_j))}
	     \sh^{\bqq_j^- - 1} (\om_j + \h)
	     \sh^{m - \bqq_j^-} (\om_j) \biggr]
	     \notag \\ & \mspace{-36.0mu}
	\det \biggl[
	   - \frac{\6^{(k-1)}_\x G(\om_j, \x)|_{\x = 0}}{(k - 1)!}
	     \biggr]
        \frac{1}{\prod_{1 \le j < k \le m} \sh( \om_j - \om_k - \h)}
	\epp
\end{align}
Here $G(\om,\x)$ is defined by (\ref{defg}) with $\fa_N$ replaced by
$\fa$.
\end{corollary}

In the following we shall give a proof of theorem \ref{theo:base}.
The proof is based on combining the two following lemmata. Lemma~%
\ref{lemma:leftact} is new and of its own interest. It describes in a
compact form the general left action of a string of monodromy matrix
elements $T^\a_\be (\x)$ conserving the $z$-component of the total spin
on a state of the form $\<0|C(\la_1) \dots C(\la_M)$. Lemma~%
\ref{lemma:matel} is a simple corollary of a result originally obtained
in \cite{GKS04a}.
\begin{lemma} \label{lemma:leftact}
The general left action. For $m, M \in {\mathbb N}$ choose $\la_1,
\dots, \la_{M + m} \in {\mathbb C}$ mutually distinct. Define
\begin{equation} \label{defam1}
     \fb_{M} (\la) = \frac{d(\la)}{a(\la)} \prod_{k = 1}^M
                      \frac{\sh(\la - \la_k + \h)}{\sh(\la - \la_k - \h)}
\end{equation}
and for $j = 1, \dots, m$ and $\ell_j \in \{1, \dots, M + m\}$
\begin{equation}
     |\ell^+| = \card \{1, \dots, M\} \cap \{\ell_j\}_{j=1}^m
\end{equation}
and $\x_j = \la_{M + j}$. Then
\begin{align} \label{gle}
     \<0| & C(\la_1) \dots C(\la_M) T^{\a_1}_{\be_1} (\x_1) \dots
            T^{\a_m}_{\be_m} (\x_m) \notag \\
        & = \sum_{\substack{\ell_1, \dots, \ell_m = 1\\
	             \ell_j \ne \ell_k\ \text{for}\ j \ne k}}^{M+m}
            \bigl( - \sh(\h) \bigr)^{|\ell^+|}
	    \biggl[ \prod_{j = |\a^+| + 1}^m \fb_M (\la_{\ell_j}) \biggr]
	    \biggl[ \prod_{j = 1}^m a (\la_{\ell_j})
	            \prod_{\substack{k = 1\\ k \ne \ell_j}}^M
		    \frac{1}{b(\la_k, \la_{\ell_j})} \biggr]
		    \notag \\ & \qqd
	    \biggl[ \prod_{1 \le j < k \le m}
	            b(\la_{\ell_k}, \la_{\ell_j}) \biggr]
	    \biggl[ \prod_{j=1}^m
	            \prod_{\substack{k = 1\\ k + M \ne \ell_j}}^m
		    \frac{1}{\sh(\la_{\ell_j} - \x_k)} \biggr]
		    \notag \\ & \qqd
	    \biggl[ \prod_{j=1}^{|\a^+|}
	            \prod_{k=1}^{\aqq_j^+ - 1}
		    \sh(\la_{\ell_j} - \x_k - \h)
	            \prod_{k = \aqq_j^+ + 1}^m
		    \sh(\la_{\ell_j} - \x_k) \biggr]
	            \notag \\ & \qqd
            \biggl[ \prod_{j = |\a^+| + 1}^{m}
	            \prod_{k=1}^{\bqq_j^- - 1}
		    \sh(\la_{\ell_j} - \x_k + \h)
	            \prod_{k = \bqq_j^- + 1}^m
		    \sh(\la_{\ell_j} - \x_k) \biggr]
            \<0| \prod_{\substack{k = 1\\
	                          k \ne \ell_1 \dots \ell_m}}^{M+m}
	                          C(\la_k) \epp
\end{align}
\end{lemma}
A proof of this lemma is given in appendix \ref{app:leftact}. Note that
the Bethe ansatz equations are not used in the derivation of (\ref{gle}).
However, (\ref{gle}) also applies if $\<0| C(\la_1) \dots C(\la_M)$
is the dominant (left) eigenvector of the quantum transfer matrix.
Then necessarily $M = N/2$ and $\fb_M = \fa_N$. This is the case needed
in the proof of theorem \ref{theo:base}.

After inserting (\ref{gle}) into the right hand side of (\ref{defdn})
it reduces to a sum over matrix elements of the form
\begin{equation}
     \<\{\x^+\} \cup \{\la^-\}|\{\la\}\>
        = \<0| \Bigl[\prod_{\x^+ \in \{\x^+\}} C(\x^+)\Bigr]
               \Bigl[\prod_{\la^- \in \{\la^-\}} C(\la^-)\Bigr]
               \Bigl[\prod_{\la \in \{\la\}} B(\la)\Bigr] |0\> \epc
\end{equation}
where $\{\x^+\} \subset \{\x\}$, $\{\la^-\} \subset \{\la\}$ and
$\card \{\x^+\} \cup \{\la^-\} = \card \{\la\} = N/2$, i.e.\
we have to deal with partitions of sets $\{X\}$ into disjoint subsets,
for which we shall adopt the general notation $p_n \{X\}$ with $n$
denoting the number of subsets. We further frequently use the shorthand
notation $|X| = \card X$ for the number of elements in a set $\{X\}$.

\begin{lemma} \label{lemma:matel}
The matrix element formula.
Let $\{\la\}$ be the set of Bethe roots that para\-meterize the dominant
eigenvalue of the quantum transfer matrix, and let $\{\x\}$ be a set
of mutually distinct complex numbers from inside the contour ${\cal C}$
such that $\{\x\} \cap \{\la\} = \emptyset$. Then, for $(\{\x^+\},
\{\x^-\}) \in p_2 \{\x\}$ and $(\{\la^+\},\{\la^-\}) \in p_2 \{\la\}$
with $|\la^+| = |\x^+|$,
\begin{multline} \label{ratio}
     \frac{\<\{\x^+\} \cup \{\la^-\}|\{\la\}\>}
          {\<\{\la\}|\{\la\}\>\prod_{j=1}^m \La_0 (\x_j)} = \\
	  \Biggl[ \prod_{j=1}^{|\x^-|}
	          \frac{\prod_{k = 1}^M b(\la_k, \x_j^-)}
		       {a(\x_j^-)(1 + \fa_N (\x_j^-))} \Biggr]
          \Biggl[ \prod_{j=1}^{|\la^+|}
	          \frac{\prod_{k=1, \la_k \ne \la_j^+}^M
		        b(\la_k, \la_j^+)}
		       {a(\la_j^+) \fa_N' (\la_j^+)} \Biggr]
	  \frac{\det G(\la_j^+,\x_k^+)}
	       {\det \Bigl[\frac{c(\la_j^+, \x_k^+)}
	                        {b(\la_j^+, \x_k^+)}\Bigr]} \epc
\end{multline}
where $M = N/2$ and where the prime denotes the derivative with respect
to the argument.
\end{lemma}
This lemma follows by applying elementary manipulations to the formula
exposed in lemma 1 of \cite{GKS05}.

Let us now proceed with the proof of theorem \ref{theo:base}. In order
to be able to combine (\ref{gle}) and (\ref{ratio}) into the right
hand side of (\ref{defdn}) we must distinguish between Bethe roots and
inhomogeneities in the sum in (\ref{gle}), i.e.\ we have to split it
into contributions with summation indices running from 1 to $M = N/2$ and
from $M + 1$ to $M + m$, respectively. Then
\begin{equation} \label{splitsum}
     \sum_{\substack{\ell_1, \dots, \ell_m = 1\\
                     \ell_j \ne \ell_k\ \text{for}\ j \ne k}}^{M+m}
		     \dots \qd =
     \sum_{(\{\e^+\},\{\e^-\}) \in p_2 ({\mathbb Z}_m)} \qd
     \sum_{\substack{\ell_{\e_1^+}, \dots, \ell_{\e_{m-n}^+} = 1\\
                     \ell_{\e_j^+} \ne \ell_{\e_k^+}\
		     \text{for}\ j \ne k}}^M \qd
     \sum_{\substack{\ell_{\e_1^-}, \dots, \ell_{\e_n^-} = M + 1\\
                     \ell_{\e_j^-} \ne \ell_{\e_k^-}\
		     \text{for}\ j \ne k}}^{M+m} \dots \epp
\end{equation}
Here ${\mathbb Z}_m = \{1, \dots, m\}$ and $n = \card \{\e^-\}$ by
definition. We shall employ the convention to enumerate the elements in
$\{\e^\pm\}$ in increasing order,
\begin{equation}
     \e_1^+ < \e_2^+ < \dots < \e_{m-n}^+ \epc \qd
     \e_1^- < \e_2^- < \dots < \e_n^- \epp
\end{equation}
Taking into account that $\ell_{\e_j^+} \in \{1, \dots, M\}$ and
$\ell_{\e_j^-} \in \{M+1, \dots, M+m\}$ we see that we must set
\begin{equation}
     \{\la^+\} = \{\la_{\ell_{\e_j^+}}\}_{j=1}^{m-n} \epc \qd
     \{\x^-\} = \{\la_{\ell_{\e_j^-}}\}_{j=1}^n \epp
\end{equation}
We enumerate the elements in $\{\la^+\}$ and $\{\x^-\}$ in such a way
that
\begin{equation} \label{enumxmlp}
     \la_j^+ = \la_{\ell_{\e_j^+}} \epc \qd
     \x_j^- = \la_{\ell_{\e_j^-}} \epp
\end{equation}
The enumeration of the elements in $\{\x^+\}$ and $\{\la^-\}$ will be
fixed later, when it is needed. Combining now (\ref{defdn}), (\ref{gle})
and (\ref{ratio}) and using (\ref{splitsum}), (\ref{enumxmlp}) we
obtain after some trivial cancellations
\begin{align} \label{proof1}
     & \mspace{-1.mu} {D^{(N)}}^{\a_1 \dots \a_m}_{\be_1 \dots \be_m}
        (\x_1, \dots, \x_m) \notag \\ & =
     \sum_{(\{\e^-\},\{\e^+\}) \in p_2 ({\mathbb Z}_m)} \qd
     \sum_{\substack{\ell_{\e_1^+}, \dots, \ell_{\e_{m-n}^+} = 1\\
                     \ell_{\e_j^+} \ne \ell_{\e_k^+}\
		     \text{for}\ j \ne k}}^M \qd
     \sum_{\substack{\ell_{\e_1^-}, \dots, \ell_{\e_n^-} = M + 1\\
                     \ell_{\e_j^-} \ne \ell_{\e_k^-}\
		     \text{for}\ j \ne k}}^{M+m}
     \det[ - G(\la_{\ell_{\e_j^+}}, \x_k^+)] \notag \\ &
     \Biggl[ \prod_{\substack{j = 1\\ \e_j^+ \le |\a^+|}}^{m-n}
	     \mspace{-8.mu}
             \frac{F_{\e_j^+} (\la_{\ell_{\e_j^+}})}
                  {\fa_N' (\la_{\ell_{\e_j^+}})} \Biggr]
     \Biggl[ \prod_{\substack{j = 1\\ \e_j^+ > |\a^+|}}^{m-n}
	     \mspace{-8.mu}
             \frac{- \overline{F}_{\e_j^+} (\la_{\ell_{\e_j^+}})}
                  {\fa_N' (\la_{\ell_{\e_j^+}})} \Biggr]
     \Biggl[ \prod_{\substack{j = 1\\ \e_j^- \le |\a^+|}}^n
	     \mspace{-8.mu}
             \frac{F_{\e_j^-} (\la_{\ell_{\e_j^-}})}
                  {1 + \fa_N (\la_{\ell_{\e_j^-}})} \Biggr]
     \Biggl[ \prod_{\substack{j = 1\\ \e_j^- > |\a^+|}}^n
	     \mspace{-8.mu}
             \frac{\overline{F}_{\e_j^-} (\la_{\ell_{\e_j^-}})}
                  {1 + \faq_N (\la_{\ell_{\e_j^-}})} \Biggr]
		  \notag \\ &
     \Biggl[ \prod_{1 \le j < k \le m} b(\la_{\ell_k}, \la_{\ell_j})
             \Biggr]
     \Biggl[ \prod_{j=1}^m \prod_{\substack{k = 1\\ k + M \ne \ell_j}}^m
             \frac{1}{\sh(\la_{\ell_j} - \x_k)} \Biggr]
     \frac{1}{\det \Bigl[
              \frac{1}{\sh(\la_{\ell_{\e_j^+}} - \x_k^+)}\Bigr]} \epp
\end{align}
Here we have used the Bethe ansatz equations in the form
$\fa_N (\la_{\ell_{\e_j^+}}) = - 1$, and we have introduced the shorthand
notations
\begin{subequations}
\label{defffbar}
\begin{align}
     F_j (\om) & = \prod_{k=1}^{\aqq_j^+ - 1} \sh(\om - \x_k - \h)
                   \prod_{k = \aqq_j^+ + 1}^m \sh(\om - \x_k) \epc
		   \qd j = 1, \dots, |\a^+| \epc \\ 
     \overline{F}_j (\om) & =
                   \prod_{k=1}^{\bqq_j^- - 1} \sh(\om - \x_k + \h)
		   \prod_{k = \bqq_j^- + 1}^m \sh(\om - \x_k) \epc
		   \qd j = |\a^+| + 1, \dots, m \epp
\end{align}
\end{subequations}
The last two terms in (\ref{proof1}) can be further simplified.
We first rewrite them as
\begin{align} \label{termv}
     & \Biggl[ \prod_{j=1}^m
               \prod_{\substack{k = 1\\ k + M \ne \ell_j}}^m
               \frac{1}{\sh(\la_{\ell_j} - \x_k)} \Biggr]
     \frac{1}{\det \Bigl[
              \frac{1}{\sh(\la_{\ell_{\e_j^+}} - \x_k^+)}\Bigr]}
	      \notag = \\ &
     \Biggl[ \prod_{1 \le j < k \le m - n}
             \frac{1}{\sh(\la_{\ell_{\e_k^+}} - \la_{\ell_{\e_j^+}})}
             \Biggr]
     \Biggl[ \prod_{j=1}^{m-n} \prod_{k=1}^n
             \frac{1}{\sh(\la_{\ell_{\e_j^+}} - \la_{\ell_{\e_k^-}})}
	     \Biggr]
     \Biggl[ \prod_{1 \le j < k \le n}
             \frac{1}{\sh(\la_{\ell_{\e_k^-}} - \la_{\ell_{\e_j^-}})}
             \Biggr]
	      \notag \\ &
     \Biggl[ \prod_{1 \le j < k \le m - n}
             \frac{1}{\sh(\x_j^+ - \x_k^+)} \Biggr]
     \Biggl[ \prod_{j=1}^n \prod_{k=1}^{m-n}
             \frac{1}{\sh(\x_j^- - \x_k^+)} \Biggr]
     \Biggl[ \prod_{1 \le j < k \le n}
             \frac{1}{\sh(\x_j^- - \x_k^-)} \Biggr] \epp
\end{align}
Then two permutations $P, Q \in {\mathfrak S}^m$ exist, such that
\begin{subequations}
\label{halfdefpq}
\begin{align}
     \e_j^- & = Pj \epc \qd j = 1, \dots, n \epc \qd
     & \e_j^+ & = P(n + j) \epc \qd j = 1, \dots, m - n \epc \\
     \label{halfdefq}
     \x_j^- & = \x_{Qj} \epc \qd j = 1, \dots, n \epc \qd
     & \x_j^+ & = \x_{Q(n + j)} \epc \qd j = 1, \dots, m - n \epp
\end{align}
\end{subequations}
After inserting (\ref{halfdefpq}) into the right hand side of
(\ref{termv}) and shifting the multiplication indices the products
can be combined into
\begin{multline}
     \Biggl[ \prod_{j=1}^m
             \prod_{\substack{k = 1\\ k + M \ne \ell_j}}^m
             \frac{1}{\sh(\la_{\ell_j} - \x_k)} \Biggr]
     \frac{1}{\det \Bigl[
              \frac{1}{\sh(\la_{\ell_{\e_j^+}} - \x_k^+)}\Bigr]} = \!
     \prod_{1 \le j < k \le m}
        \frac{1}{\sh(\la_{\ell_{Pk}} - \la_{\ell_{Pj}})
                 \sh(\x_{Qj} - \x_{Qk})} \\ =
     \sign(PQ) \prod_{1 \le j < k \le m}
               \frac{1}{\sh(\la_{\ell_k} - \la_{\ell_j})
	                \sh(\x_j - \x_k)} \epp
\end{multline}
Inserting this back into (\ref{proof1}) we conclude that
\begin{align} \label{proof2}
     & \mspace{-1.mu} {D^{(N)}}^{\a_1 \dots \a_m}_{\be_1 \dots \be_m}
        (\x_1, \dots, \x_m)
	\prod_{1 \le j < k \le m} \sh(\x_j - \x_k) \notag \\ & =
     \sum_{(\{\e^+\},\{\e^-\}) \in p_2 ({\mathbb Z}_m)} \qd
     \sum_{\ell_{\e_1^+}, \dots, \ell_{\e_{m-n}^+} = 1}^M \qd
     \sum_{\substack{\ell_{\e_1^-}, \dots, \ell_{\e_n^-} = M + 1\\
                     \ell_{\e_j^-} \ne \ell_{\e_k^-}\
		     \text{for}\ j \ne k}}^{M+m}
     \frac{\sign(PQ) \det[ - G(\la_{\ell_{\e_j^+}}, \x_k^+)]}
	            {\prod_{1 \le j < k \le m}
		     \sh(\la_{\ell_j} - \la_{\ell_k} - \h)} \notag \\ &
     \Biggl[ \prod_{\substack{j = 1\\ \e_j^+ \le |\a^+|}}^{m-n}
	     \mspace{-10.mu}
             \frac{F_{\e_j^+} (\la_{\ell_{\e_j^+}})}
                  {\fa_N' (\la_{\ell_{\e_j^+}})} \Biggr]
     \Biggl[ \prod_{\substack{j = 1\\ \e_j^+ > |\a^+|}}^{m-n}
	     \mspace{-10.mu}
             \frac{- \overline{F}_{\e_j^+} (\la_{\ell_{\e_j^+}})}
                  {\fa_N' (\la_{\ell_{\e_j^+}})} \Biggr]
     \Biggl[ \prod_{\substack{j = 1\\ \e_j^- \le |\a^+|}}^n
	     \mspace{-10.mu}
             \frac{F_{\e_j^-} (\la_{\ell_{\e_j^-}})}
                  {1 + \fa_N (\la_{\ell_{\e_j^-}})} \Biggr]
     \Biggl[ \prod_{\substack{j = 1\\ \e_j^- > |\a^+|}}^n
	     \mspace{-10.mu}
             \frac{\overline{F}_{\e_j^-} (\la_{\ell_{\e_j^-}})}
                  {1 + \faq_N (\la_{\ell_{\e_j^-}})} \Biggr] \epp
\end{align}
Note that we could leave out the exclusions in the sums over
$\ell_{\e_j^+}, j = 1, \dots, m - n$, since
$\det[ - G(\la_{\ell_{\e_j^+}}, \x_k^+)]$ vanishes if two of the
$\ell_{\e_j^+}$ coincide. Note further that $Q$ is not entirely defined
by (\ref{halfdefpq}), since so far we did not fix a convention for
enumerating the $\x_j^+$. It follows from (\ref{enumxmlp}) and
(\ref{halfdefq}) that
\begin{equation} \label{qxi}
     \x_j^- = \la_{\ell_{\e_j^-}} = \x_{Qj} = \x_{\ell_{\e_j^-} - M}
        \qd \Leftrightarrow \qd Qj = \ell_{\e_j^-} - M \epc \qd
	j = 1, \dots, n \epp
\end{equation}
In order to fix $Q$ uniquely we stipulate that
\begin{equation} \label{fixq}
     Q(n + 1) < Q(n + 2) < \dots < Qm \epp
\end{equation}
This then also fixes the enumeration of elements in $\{\x^+\}$ through
(\ref{halfdefq}). Our freedom in the choice of the enumeration is
explained by the fact that $\sign (Q)$ is multiplied by
$\det[ - G(\la_{\ell_{\e_j^+}}, \x_k^+)]$. The product is totally
symmetric in the $\x_j^+$.

In the next step we would like to make the dependence of the terms
under the sum in (\ref{proof2}) on the inhomogeneities more explicit
and rewrite the sum over the $\ell_{\e_j^-}$ in a more convenient form.
The inhomogeneities occur only in the last two products, where we
can directly use (\ref{qxi}), and in the product in the denominator,
which we shall write as
\begin{equation} \label{ximinprod}
     \prod_{1 \le j < k \le m} \sh(\la_{\ell_j} - \la_{\ell_k} - \h)
        = \prod_{1 \le j < k \le m} \sh(\om_j - \om_k - \h)
	  \Bigr|_{\om_{\e_j^+} = \la_{\ell_{\e_j^+}}, \,
	                    \om_{\e_j^-} = \x_{\ell_{\e_j^- - M}}} \epp
\end{equation}

In the multiple sum over $\ell_{\e_1^-}, \dots, \ell_{\e_n^-}$ all
summation variables must take mutually distinct values. For this reason
each term in the multiple sum defines a partition $(\{\de^+\},
\{\de^-\}) \in p_2 ({\mathbb Z}_m)$ via
\begin{equation}
     \{\de^-\} = \{\ell_{\e_1^-} - M, \dots, \ell_{\e_n^-} - M\} \epc
        \qd \{\de^+\} = {\mathbb Z}_m \setminus \{\de^-\} \epp
\end{equation}
By our convention the elements in $\{\de^\pm\}$ are ordered, $\de_j^\pm <
\de_k^\pm$ if $j < k$. Hence, a unique $R \in {\mathfrak S}^n$ exists,
such that
\begin{equation} \label{defpermr}
     \de_{R j}^- = \ell_{\e_j^-} - M \epc \qd j = 1, \dots, n \epp
\end{equation}
On the other hand, each pair consisting of a partition $(\{\de^+\},
\{\de^-\}) \in p_2 ({\mathbb Z}_m)$ with $|\de^-| = n$ and a permutation
$R \in {\mathfrak S}^n$ uniquely determines  an $n$-tupel
$(\ell_{\e_1^-}, \dots, \ell_{\e_n^-})$ of summation variables by
equation (\ref{defpermr}). We conclude that with the substitution
(\ref{defpermr}) the sum over the $\ell_{\e_j^-}$ in (\ref{proof2}) may
be split as
\begin{equation} \label{splitsum2}
     \sum_{\substack{\ell_{\e_1^-}, \dots, \ell_{\e_n^-} = M + 1\\
                     \ell_{\e_j^-} \ne \ell_{\e_k^-}\
		     \text{for}\ j \ne k}}^{M+m} \dots \qd =
     \sum_{\substack{(\{\de^+\}, \{\de^-\}) \in p_2 ({\mathbb Z}_m) \\
                     |\de^-| = n}} \qd
     \sum_{R \in {\mathfrak S}^n} \dots \epp
\end{equation}

The permutation $R \in {\mathfrak S}^n$ has a natural embedding into
${\mathfrak S}^m$ defined by
\begin{equation} \label{defpermrtilde}
     \tilde R = \begin{cases} Rj & \text{if $j = 1, \dots, n$} \\
                               j & \text{if $j = n + 1, \dots, m$.}
                \end{cases}
\end{equation}
We would like to split off $\tilde R$ from the permutation $Q$ defined
in (\ref{halfdefq}), (\ref{fixq}). Using (\ref{qxi}), (\ref{defpermr})
and (\ref{defpermrtilde}) we obtain
\begin{equation}
     Qj = \ell_{\e_j^-} - M = \de_{\tilde R j}^- \qd \Leftrightarrow \qd
     \de_j^- = Q \tilde{R}^{- 1} j \epc \qd j = 1, \dots, n \epp
\end{equation}
It follows that $\tilde Q = Q \tilde{R}^{- 1}$ satisfies
\begin{equation}
     \tilde Q 1 < \tilde Q 2 < \dots < \tilde Q n \epc \qqd
     \tilde Q (n+1) < \tilde Q (n+2) < \dots < \tilde Q m \epp
\end{equation}
Thus, there is a one-to-one correspondence between the permutation
$\tilde Q$ and the partition $(\{\de^+\}, \{\de^-\}) \in p_2
({\mathbb Z}_m)$,
\begin{equation}
     \de_j^- = \tilde Q j \epc \qd j = 1, \dots, n \epc \qqd
     \de_j^+ = \tilde Q (n + j) \epc \qd j = 1, \dots, m - n \epc
\end{equation}
and $\tilde Q$ does not depend on $R$. With (\ref{halfdefq}) we find that
\begin{equation} \label{xip}
     \x_j^+ = \x_{\tilde Q (n+j)} = \x_{\de_j^+} \epp
\end{equation}

Now we are in a position to rewrite the sum (\ref{proof2}) in a form
suitable for transforming it into a multiple integral. We use
(\ref{splitsum2}) in (\ref{proof2}) and insert (\ref{qxi}),
(\ref{ximinprod}), (\ref{defpermr}) and (\ref{xip}),
\begin{multline} \label{proof3}
     {D^{(N)}}^{\a_1 \dots \a_m}_{\be_1 \dots \be_m}
        (\x_1, \dots, \x_m)
	\prod_{1 \le j < k \le m} \sh(\x_j - \x_k) \\ =
     \sum_{(\{\e^+\},\{\e^-\}) \in p_2 ({\mathbb Z}_m)} \qd
     \sum_{\ell_{\e_1^+}, \dots, \ell_{\e_{m-n}^+} = 1}^M \qd
     \sum_{\substack{(\{\de^+\}, \{\de^-\}) \in p_2 ({\mathbb Z}_m) \\
                     |\de^-| = n}} \sign(P \tilde Q) \\
     \sum_{R \in {\mathfrak S}^n}
     \frac{\sign(R) \det[ - G(\la_{\ell_{\e_j^+}}, \x_{\de_k^+})]}
	  {\prod_{1 \le j < k \le m} \sh(\om_j - \om_k - \h)
           \Bigr|_{\om_{\e_j^+} = \la_{\ell_{\e_j^+}}, \:
	           \om_{\e_j^-} = \x_{\de_{Rj}^-}}} \\ \mspace{-6.mu}
     \Biggl[ \prod_{\substack{j = 1\\ \e_j^+ \le |\a^+|}}^{m-n}
	     \mspace{-11.mu}
             \frac{F_{\e_j^+} (\la_{\ell_{\e_j^+}})}
                  {\fa_N' (\la_{\ell_{\e_j^+}})} \Biggr]
     \Biggl[ \prod_{\substack{j = 1\\ \e_j^+ > |\a^+|}}^{m-n}
	     \mspace{-11.mu}
             \frac{- \overline{F}_{\e_j^+} (\la_{\ell_{\e_j^+}})}
                  {\fa_N' (\la_{\ell_{\e_j^+}})} \Biggr]
     \Biggl[ \prod_{\substack{j = 1\\ \e_j^- \le |\a^+|}}^n
	     \mspace{-11.mu}
             \frac{F_{\e_j^-} (\x_{\de_{Rj}^-})}
                  {1 + \fa_N (\x_{\de_{Rj}^-})} \Biggr]
     \Biggl[ \prod_{\substack{j = 1\\ \e_j^- > |\a^+|}}^n
	     \mspace{-11.mu}
             \frac{\overline{F}_{\e_j^-} (\x_{\de_{Rj}^-})}
                  {1 + \faq_N (\x_{\de_{Rj}^-})} \Biggr] \epp
\end{multline}

Recall that we have assumed that the inhomogeneities $\x_j$ are mutually
distinct and distinct from the Bethe roots. Hence, our canonical contour
${\cal C}$ can be decomposed into two simple contours ${\cal B}$ and
${\cal I}$, such that
\begin{equation}
     {\cal C = B + I} \epc
\end{equation}
and ${\cal B}$ contains only Bethe roots and no inhomogeneities, while
${\cal I}$ contains only inhomogeneities but no Bethe roots. Then
$1/(1 + \fa_N (\om))$ is holomorphic inside ${\cal I}$ and meromorphic
inside ${\cal B}$ with only simple poles, located at the Bethe roots
\cite{GKS04a}. On the other hand, $- G(\om,\x_j)$ considered as a
function of $\om$ is holomorphic inside ${\cal B}$ and meromorphic
inside ${\cal I}$ with a single simple pole with residue $1$ at $\x_j$
\cite{GKS04a}. Thus,
\begin{align} \label{proof4}
     & {D^{(N)}}^{\a_1 \dots \a_m}_{\be_1 \dots \be_m}
        (\x_1, \dots, \x_m)
	\prod_{1 \le j < k \le m} \sh(\x_j - \x_k)
	\displaybreak[0] \notag \\ & =
     \sum_{(\{\e^+\},\{\e^-\}) \in p_2 ({\mathbb Z}_m)} \qd
     \sum_{\ell_{\e_1^+}, \dots, \ell_{\e_{m-n}^+} = 1}^M \qd
     \sum_{\substack{(\{\de^+\}, \{\de^-\}) \in p_2 ({\mathbb Z}_m) \\
                     |\de^-| = n}} \sign(P \tilde Q) \:
     \biggl[ \prod_{j=1}^n \int_{\cal I} \frac{\rd \om_{\e_j^-}}{2 \p \i}
             \biggr] \notag \\ &
     \underbrace{\sum_{R \in {\mathfrak S}^n} \sign(R)
        \bigl[- G(\om_{\e_1^-}, \x_{\de_{R1}^-}) \bigr] \dots
        \bigl[- G(\om_{\e_n^-}, \x_{\de_{Rn}^-}) \bigr]}_{%
	      \det[ - G(\om_{\e_j^-}, \x_{\de_k^-})]}
     \frac{\det[ - G(\la_{\ell_{\e_j^+}}, \x_{\de_k^+})]}
	  {\dst{\prod_{1 \le j < k \le m}} \sh(\om_j - \om_k - \h)
           \Bigr|_{\om_{\e_j^+} = \la_{\ell_{\e_j^+}}}}
	   \notag \displaybreak[0] \\ & \mspace{-6.mu}
     \Biggl[ \prod_{\substack{j = 1\\ \e_j^+ \le |\a^+|}}^{m-n}
	     \mspace{-11.mu}
             \frac{F_{\e_j^+} (\la_{\ell_{\e_j^+}})}
                  {\fa_N' (\la_{\ell_{\e_j^+}})} \Biggr]
     \Biggl[ \prod_{\substack{j = 1\\ \e_j^+ > |\a^+|}}^{m-n}
	     \mspace{-11.mu}
             \frac{- \overline{F}_{\e_j^+} (\la_{\ell_{\e_j^+}})}
                  {\fa_N' (\la_{\ell_{\e_j^+}})} \Biggr]
     \Biggl[ \prod_{\substack{j = 1\\ \e_j^- \le |\a^+|}}^n
	     \mspace{-11.mu}
             \frac{F_{\e_j^-} (\om_{\e_j^-})}
                  {1 + \fa_N (\om_{\e_j^-})} \Biggr]
     \Biggl[ \prod_{\substack{j = 1\\ \e_j^- > |\a^+|}}^n
	     \mspace{-11.mu}
             \frac{\overline{F}_{\e_j^-} (\om_{\e_j^-})}
                  {1 + \faq_N (\om_{\e_j^-})} \Biggr] 
		  \displaybreak[0] \notag \\[3ex] & =
     \underbrace{
     \sum_{(\{\e^+\},\{\e^-\}) \in p_2 ({\mathbb Z}_m)} \qd
     \biggl[ \prod_{j=1}^{m-n}
             \int_{\cal B} \frac{\rd \om_{\e_j^+}}{2 \p \i} \biggr]
     \biggl[ \prod_{j=1}^n
             \int_{\cal I} \frac{\rd \om_{\e_j^-}}{2 \p \i} \biggr]}_{%
        = \prod_{j=1}^m \int_{\cal C} \frac{\rd \om_j}{2 \p \i}} 
     \frac{1}{\prod_{1 \le j < k \le m} \sh(\om_j - \om_k - \h)}
     \notag \\[1ex] & \underbrace{
     \sum_{\substack{(\{\de^+\}, \{\de^-\}) \in p_2 ({\mathbb Z}_m) \\
                     |\de^-| = n}} \sign(P \tilde Q)
     \det[ - G(\om_{\e_j^-}, \x_{\de_k^-})]
     \det[ - G(\om_{\e_j^+}, \x_{\de_k^+})]}_{%
        = \det[ - G(\om_j, \x_k)]} \notag \\ & \mspace{-6.mu}
     \Biggl[ \prod_{\substack{j = 1\\ \e_j^+ \le |\a^+|}}^{m-n}
	     \mspace{-11.mu}
             \frac{F_{\e_j^+} (\om_{\e_j^+})}
                  {1 + \fa_N (\om_{\e_j^+})} \Biggr]
     \Biggl[ \prod_{\substack{j = 1\\ \e_j^+ > |\a^+|}}^{m-n}
	     \mspace{-11.mu}
             \frac{\overline{F}_{\e_j^+} (\om_{\e_j^+})}
                  {1 + \faq_N (\om_{\e_j^+})} \Biggr]
     \Biggl[ \prod_{\substack{j = 1\\ \e_j^- \le |\a^+|}}^n
	     \mspace{-11.mu}
             \frac{F_{\e_j^-} (\om_{\e_j^-})}
                  {1 + \fa_N (\om_{\e_j^-})} \Biggr]
     \Biggl[ \prod_{\substack{j = 1\\ \e_j^- > |\a^+|}}^n
	     \mspace{-11.mu}
             \frac{\overline{F}_{\e_j^-} (\om_{\e_j^-})}
                  {1 + \faq_N (\om_{\e_j^-})} \Biggr] \notag \\[3ex] &
     \mspace{-6.mu}
     = \biggl[ \prod_{j=1}^{|\a^+|}
               \int_{\cal C} \frac{\rd \om_j}{2 \p \i}
               \frac{F_j (\om_j)}{1 + \fa_N (\om_j)} \biggr]
	       \mspace{-3.mu}
       \biggl[ \prod_{j = |\a^+| + 1}^m
               \int_{\cal C} \frac{\rd \om_j}{2 \p \i}
               \frac{\overline{F}_j (\om_j)}{1 + \faq_N (\om_j)} \biggr]
       \frac{\det[ - G(\om_j, \x_k)]}
            {\dst{\prod_{1 \le j < k \le m}} \sh(\om_j - \om_k - \h)}
	     \epc
\end{align}
and our proof of theorem \ref{theo:base} is complete. Note that we used
the Laplace expansion for determinants in the third equation.

\section{The two-point functions}
In this section we shall derive a number of formulae for two-point
correlation functions. Since the derivations concern only the `algebraic
part' of the integrals, namely the product $1/\prod_{1 \le j < k \le m}
\sh(\om_j - \om_k - \h)$ and the functions $F_j$, $\overline{F}_j$
introduced in equations (\ref{defffbar}), we may strongly refer to the
zero temperature results obtained in \cite{KMST04c}.

Let us start with a certain one-parameter generating function of the
$S^z$-$S^z$ correlation functions that was introduced in \cite{IzKo84}.
It is defined as the thermal average
\begin{equation} \label{defgenfun}
     \PH (\ph|m) = \bigl\< \exp \{\ph \sum_{j=1}^m {e_j}_2^2 \}
                           \bigr\>_{T, h} \epp
\end{equation}
The function $\PH (\ph|m)$ generates the two-point function through
the action of a second order differential-difference operator,
\begin{equation} \label{genfunappl}
     \bigl\< S_1^z S_m^z \bigr\>_{T, h} = \tst{\frac{1}{4}}
          (2 D_m^2 \6_\ph^2 - 4 D_m \6_\ph + 1) 
                \PH (\ph|m) \Bigr|_{\ph = 0} \epp
\end{equation}
Here $D_m$ denotes the `lattice derivative' defined on any sequence
$(z_n)_{n \in {\mathbb N}}$ of complex numbers by $D_m z_m = z_m -
z_{m-1}$.

The function $\PH (\ph| m)$ is particularly useful and comparatively
simple, since it is closely connected to a twisted version of the
quantum transfer matrix as can be seen by inserting $X^{(n)} =
\bigl( \begin{smallmatrix} 1 & 0 \\ 0 & \re^\ph \end{smallmatrix} \bigr)$
for $n = 1, \dots, m$ into our basic formula (\ref{cor2}). Then
\begin{equation}
     \PH (\ph| m) = \lim_{N \rightarrow \infty} \:
                    \lim_{\x_1, \dots, \x_m \rightarrow 0}
                    \PH^{(N)} (\ph| \x_1, \dots, \x_m) \epc
\end{equation}
where
\begin{equation}
     \PH^{(N)} (\ph| \x_1, \dots, \x_m)
        = \frac{\< \{\la\}| \prod_{j=1}^m
	                    \bigl( A (\x_j) + \re^\ph D (\x_j) \bigr)
			    |\{\la\}\>}
	       {\<\{\la\}|\{\la\}\> \prod_{j=1}^m \La_0 (\x_j)} \epp
\end{equation}
is a polynomial of degree $m$ in $\re^\ph$.

We see from (\ref{densint}) that the Trotter limit and the homogeneous
limit $\x_j \rightarrow 0$ commute. Hence, it makes sense to perform
the Trotter limit in (\ref{densint}) and to consider `inhomogeneous
generalizations of correlation functions'. An example for such an
inhomogeneous generalization is the density matrix element itself,
\begin{equation}
     D^{\a_1 \dots \a_m}_{\be_1 \dots \be_m} (\x_1, \dots, \x_m) =
        \lim_{N \rightarrow \infty}
	{D^{(N)}}^{\a_1 \dots \a_m}_{\be_1 \dots \be_m}
	          (\x_1, \dots, \x_m) \epc
\end{equation}
obtained from (\ref{densint}) by taking the Trotter limit, i.e., by
replacing $\fa_N$ by $\fa$.

Let us define
\begin{equation} \label{fsumd}
     f_n (\x_1, \dots, \x_m) =
        \sum_{\substack{\a_1, \dots, \a_m = 1, 2\\
	                \a_1 + \dots + \a_m = 2m - n}}
	D^{\a_1 \dots \a_m}_{\a_1 \dots \a_m} (\x_1, \dots, \x_m) \epp
\end{equation}
Then
\begin{equation}
     \PH (\ph| \x_1, \dots, \x_m) =
        \lim_{N \rightarrow \infty} \PH^{(N)} (\ph| \x_1, \dots, \x_m)
	= \sum_{n=0}^m \re^{(m-n) \ph} f_n (\x_1, \dots, \x_m)
\end{equation}
is the inhomogeneous generalization of $\PH (\ph|m)$, equation
(\ref{defgenfun}). Following \cite{KMST04c} we shall derive an integral
representation for the coefficients $f_n (\x_1, \dots, \x_m)$ of the
polynomial $\PH (\ph| \x_1, \dots, \x_m)$. This gives an integral
formula for the generating function (\ref{defgenfun}) in the homogeneous
limit. The coefficients $f_n$ (in the homogeneous limit) have a simple
physical interpretation. They represent the probability for a segment
of length $m$ of the infinite chain to be in a state with $z$-component
of the total spin equal to $n - m/2$. For instance, for $n = m$ we find
$f_m = \lim_{\ph \rightarrow - \infty} \PH (\ph|m)$ which is the
probability to have maximal $S^z = m/2$ and which is, of course,
equal to the emptiness formation probability for spin-up. Thus, one may
think of the $f_n$ as generalizations of the emptiness formation
probability.
\begin{theorem} \label{theo:coefff}
Define
\begin{equation}
     t(\om, \x) = \frac{\sh(\h)}{\sh(\om - \x) \sh(\om - \x + \h)}
\end{equation}
and for $n = 0, \dots, m$
\begin{equation}
     M_n (\om_j, \x_k) = \begin{cases}
                         t(\x_k, \om_j) & \text{for $1 \le j \le n$,} \\
                         t(\om_j, \x_k) & \text{for $n < j \le m$.}
                         \end{cases}
\end{equation}
Then the following integral formula holds for the coefficient $f_n$
in the expansion of the inhomogeneous generalization of the 
generating function of the $S^z$-$S^z$ correlation functions with
respect to $\re^\ph$,
\begin{multline} \label{gencoeffint}
     f_n (\x_1, \dots, \x_m) = \frac{(-1)^n}{n! (m - n)!}
	  \biggl[ \prod_{j=1}^n
	     \int_{\cal C} \frac{d \om_j}{2 \p \i (1 + \fa (\om_j))}
	     \biggr]
          \biggl[ \prod_{j = n + 1}^{m}
	     \int_{\cal C} \frac{d \om_j}{2 \p \i (1 + \faq (\om_j))}
	     \biggr] \\[1ex]
	  \biggl[ \prod_{k=1}^m \prod_{j=1}^n
	     \frac{\sh(\om_j - \x_k - \h) \sh(\om_j - \x_k)}
	          {\sh(\om_j - \om_k - \h)}
          \prod_{j = n + 1}^{m}
	     \frac{\sh(\om_j - \x_k + \h) \sh(\om_j - \x_k)}
	          {\sh(\om_j - \om_k + \h)} \biggr] \\[1ex]
        \frac{\det( - G(\om_j, \x_k))}
	     {\prod_{1 \le j < k \le m} \sh(\x_k - \x_j)} \:
        \frac{\det(M_n (\om_j, \x_k))}
	     {\prod_{1 \le j < k \le m} \sh(\x_j - \x_k)} \epp
\end{multline}
\end{theorem}
A proof of this theorem is shown in appendix \ref{app:emptgen}.
\begin{corollary}
Integral formula for $f_n$ in the homogeneous limit.
\begin{multline} \label{geninthom}
     f_n =
        \biggl[ \prod_{j=1}^n
	     \int_{\cal C} \frac{d \om_j}{2 \p \i}
	     \frac{\sh^m (\om_j - \h) \sh^m (\om_j)}{1 + \fa (\om_j)}
	     \biggr]
        \biggl[ \prod_{j = n + 1}^m
	     \int_{\cal C} \frac{d \om_j}{2 \p \i}
	     \frac{\sh^m (\om_j + \h) \sh^m (\om_j)}{1 + \faq (\om_j)}
	     \biggr] \\[1ex]
        \frac{(-1)^{n + m(m - 1)/2}}{n! (m - n)!}
	  \biggl[ \prod_{k=1}^m \prod_{j=1}^n
	     \frac{1}{\sh(\om_j - \om_k - \h)}
          \prod_{j = n + 1}^{m}
	     \frac{1}{\sh(\om_j - \om_k + \h)} \biggr] \\[1ex]
	\det \biggl[
	   - \frac{\6^{(k-1)}_\x G(\om_j, \x)|_{\x = 0}}{(k - 1)!}
	     \biggr] \:
	\det \biggl[
	     \frac{\6^{(k-1)}_\x M_n (\om_j, \x)|_{\x = 0}}{(k - 1)!}
	     \biggr] \epp
\end{multline}
\end{corollary}
Similar but slightly more complicated formulae can be derived for the
general two-point functions for points at distance $m$ defined by
\begin{equation}
     G^{\a \g}_{\be \de} (m; T, h) =
        \lim_{\x_1, \dots, \x_{m+1} \rightarrow 0}
	G^{\a \g}_{\be \de} (\x_1, \dots, \x_{m+1}) \epc
\end{equation}
where
\begin{align} \label{ggexp}
     G^{\a \g}_{\be \de} (\x_1, \dots, \x_{m+1}) & =
        \sum_{n=0}^{m-1} {g_n}^{\a \g}_{\be \de} (\x_1, \dots, \x_{m+1})
	\epc \\
     {g_n}^{\a \g}_{\be \de} (\x_1, \dots, \x_{m+1}) & =
        \sum_{\substack{\a_2, \dots, \a_m = 1, 2\\
	                \a_2 + \dots + \a_m = 2m - n - 2}}
	D^{\a \a_2 \dots \a_m \g}_{\be \a_2 \dots \a_m \de}
	   (\x_1, \dots, \x_{m+1}) \epp
\end{align}
We shall consider
\begin{equation}
     \< \s_1^+ \s_{m+1}^- \>_{T, h} = G^{21}_{12} (m; T, h)
\end{equation}
as an example.
\begin{theorem}
The coefficients ${g_n}^{21}_{12} (\x_1, \dots, \x_{m+1})$ in the
expansion (\ref{ggexp}) of the inhomogeneous generalization of the
two-point function $\< \s_1^+ \s_{m+1}^- \>_{T, h}$ are given by the
integrals
\begin{multline}
     {g_n}^{21}_{12} (\x_1, \dots, \x_{m+1}) = \\
        \frac{(-1)^n}{n! (m - n - 1)!}
	  \biggl[ \prod_{j=1}^{n+1}
	     \int_{\cal C} \frac{d \om_j}{2 \p \i (1 + \fa (\om_j))}
	     \biggr]
          \biggl[ \prod_{j = n + 2}^{m+1}
	     \int_{\cal C} \frac{d \om_j}{2 \p \i (1 + \faq (\om_j))}
	     \biggr] \\[1ex]
        \frac{\det_{m+1} (G(\om_j, \x_k))}
	     {\prod_{1 \le j < k \le m + 1} \sh(\x_j - \x_k)} \:
        \frac{\det_{m-1} (M_n (\om_{j+1}, \x_{k+1}))}
	     {\prod_{2 \le j < k \le m} \sh(\x_k - \x_j)}
	     \frac{1}{\sh(\om_{m+1} - \om_1 + \h)} \\[1ex]
	\frac{\prod_{k=1}^m \prod_{j=1}^{n+1} \sh(\om_j - \x_k - \h)
	      \prod_{j = n + 2}^{m+1} \sh(\om_j - \x_k + \h)}
	     {\prod_{k=2}^m \prod_{j=1}^{n+1} \sh(\om_j - \om_k - \h)
	      \prod_{j = n + 2}^{m+1} \sh(\om_j - \om_k + \h)}
	\biggl[ \prod_{k=2}^{m+1} \prod_{j=2}^m \sh(\om_j - \x_k) \biggr]
	      \epp
\end{multline}
\end{theorem}
A proof of the formula is easily obtained by combining the ideas of
appendix \ref{app:emptgen} with those of \cite{KMST04c}.
\begin{corollary}
Integral formula for ${g_n}^{21}_{12}$ in the homogeneous limit.
\begin{multline} \label{pminthom}
     {g_n}^{21}_{12} = 
        \biggl[ \prod_{j=1}^{n+1}
	     \int_{\cal C} \frac{d \om_j}{2 \p \i}
	     \frac{\sh^m (\om_j - \h) \sh^m (\om_j)}{1 + \fa (\om_j)}
	     \biggr]
        \biggl[ \prod_{j = n + 2}^{m+1}
	     \int_{\cal C} \frac{d \om_j}{2 \p \i}
	     \frac{\sh^m (\om_j + \h) \sh^m (\om_j)}{1 + \faq (\om_j)}
	     \biggr] \\[1ex]
        \frac{(-1)^{n + 1 + m(m - 1)/2}}{n! (m - n - 1)!}
	  \biggl[ \prod_{k=2}^m \prod_{j=1}^{n+1}
	     \frac{1}{\sh(\om_j - \om_k - \h)}
          \prod_{j = n + 2}^{m+1}
	     \frac{1}{\sh(\om_j - \om_k + \h)} \biggr] \\[1ex]
	\tst{\det_{m+1}} \biggl[
	   - \frac{\6^{(k-1)}_\x G(\om_j, \x)|_{\x = 0}}{(k - 1)!}
	     \biggr] \:
	\tst{\det_{m-1}} \biggl[
	     \frac{\6^{(k-1)}_\x M_n (\om_{j+1}, \x)|_{\x = 0}}{(k - 1)!}
	     \biggr] \\[1ex]
	\frac{1}{\sh^m (\om_1) \sh^m (\om_{m+1})
	         \sh(\om_{m+1} - \om_1 + \h)} \epp
\end{multline}
\end{corollary}

\section{Conclusions}
We presented a proof of the integral representation of the density
matrix (\ref{densint}) and obtained efficient formulae for
a one-parameter generating function of the $S^z$-$S^z$ correlation
function and for $\< \s_1^+ \s_{m+1}^- \>_{T, h}$. The latter formulae
were obtained by applying the re-summation technique developed in
\cite{KMST04c} in the finite temperature case. This technique applies
to arbitrary finite-temperature two-point functions.

We hope that our results will prove to be as useful as the multiple
integral formulae in the zero temperature case and will eventually lead
to explicit expressions for short-range correlation functions reaching
beyond our preliminary numerical calculations of the integrals in
\cite{BoGo05} (for the most recent progress at $T = 0$ see
\cite{SaSh05,SST05pp}). We believe that the integral formulae can be
efficiently used to obtain high-order high-temperature expansions for
the two-point functions, closely following the example \cite{TsSh05}
of the emptiness formation probability. We further hope that we may be
able to extract the long-distance asymptotic behaviour of the
correlation functions directly from the integrals (for the zero
temperature case compare \cite{KMST02d,KLNS03}). These and many other
interesting questions are still open, but we may now be close to their
answers.\\[1ex]
{\bf Acknowledgement.}
The authors would like to thank H. E. Boos and A. Kl\"umper for helpful
discussions. This work was partially supported by the Deutsche
Forschungsgemeinschaft under grant number Go 825/4-2.

\clearpage

{\appendix
\Appendix{The general left action} \label{app:leftact}
\noindent
Here we shall provide a proof of lemma \ref{lemma:leftact}. This means
that we have to calculate the left action of a product of monodromy
matrix elements on a state of the form
\begin{equation} \label{cform}
     \<\{\la_j\}_{j=1}^M| = \<0| \prod_{k=1}^M C(\la_k) \epc
\end{equation}
where $\la_1, \dots, \la_M \in {\mathbb C}$ are mutually distinct. We
shall restrict ourselves to products $T^{\a_1}_{\be_1} (\x_1) \dots
T^{\a_m}_{\be_m} (\x_m)$ which contain as many $B$s as $C$s, i.e.\
products for which
\begin{equation} \label{constraintab}
     \sum_{j=1}^m \a_j = \sum_{j=1}^m \be_j \epp
\end{equation}
Only these products have non-zero expectation value $\<\{\la\}|
T^{\a_1}_{\be_1} (\x_1) \dots T^{\a_m}_{\be_m} (\x_m) |\{\la\}\>$.

In order to obtain the action of a whole product of monodromy matrix
elements we first of all need compact expressions for the action of
a single monodromy matrix element. As was observed in \cite{KMT99b}
the known expressions for the single action (see e.g.\ \cite{KBIBo})
can be written more compactly if one introduces the notation $\x_j
= \la_{M+j}$ for the inhomogeneity parameters. This idea was extended
to its full beauty in \cite{GKS05}, where the equations
\begin{align}
     \mspace{-9.mu} \label{ccca}
     \<0| & \Bigl[ \prod_{k=1}^M C(\la_k) \Bigr] A (\la_{M+1})
        = \sum_{\ell = 1}^{M+1} a(\la_\ell) c(\la_{M+1}, \la_\ell)
	  \biggr[ \prod_{\substack{k = 1 \\ k \ne \ell}}^{M+1}
	         \frac{1}{b(\la_k, \la_\ell)} \biggl]
		 \<0| \prod_{\substack{k = 1 \\ k \ne \ell}}^{M+1}
		 C(\la_k), \\[2ex]
     \mspace{-9.mu} \label{cccd}
     \<0| & \Bigl[ \prod_{k=1}^M C(\la_k) \Bigr] D (\la_{M+1})
        = \sum_{\ell = 1}^{M+1} d(\la_\ell) c(\la_\ell, \la_{M+1})
	  \biggr[ \prod_{\substack{k = 1 \\ k \ne \ell}}^{M+1}
	         \frac{1}{b(\la_\ell, \la_k)} \biggl]
		 \<0| \prod_{\substack{k = 1 \\ k \ne \ell}}^{M+1}
		 C(\la_k), \\[2ex]
     \mspace{-9.mu} \label{cccb}
     \<0| & \Bigl[ \prod_{k=1}^M C(\la_k) \Bigr] B (\la_{M+1})
        = \sum_{\ell_1 = 1}^{M+1}
	  \sum_{\substack{\ell_2 = 1 \\ \ell_2 \ne \ell_1}}^{M+1}
	  d(\la_{\ell_1}) c(\la_{\ell_1}, \la_{M+1})
	  \biggr[ \prod_{\substack{k = 1 \\ k \ne \ell_1}}^{M+1}
	         \frac{1}{b(\la_{\ell_1}, \la_k)} \biggl]
		 \notag \\ & \mspace{153.0mu}
          a(\la_{\ell_2}) c(\la_{M+1}, \la_{\ell_2})
	  \biggr[ \prod_{\substack{k = 1 \\ k \ne \ell_1, \ell_2}}^{M+1}
	         \frac{1}{b(\la_k, \la_{\ell_2})} \biggl] \<0|
		 \prod_{\substack{k = 1 \\ k \ne \ell_1, \ell_2}}^{M+1}
		 C(\la_k).
\end{align}
were obtained.

These equations are an appropriate starting point for our proof. The
natural thing to do is using them iteratively. Then each operator
$A$ or $D$ generates a single sum and each operator $B$ a double
sum. The number $n$ of these sums after acting with the first $j$
monodromy matrix elements $T^{\a_1}_{\be_1} \dots T^{\a_j}_{\be_j}$ is
equal to the number of up-spins in the subsequence $(\a_n)_{n=1}^j$
plus the number of down-spins in the subsequence $(\be_n)_{n=1}^j$.
This number may be zero if all monodromy matrix elements
$T^{\a_1}_{\be_1} \dots T^{\a_j}_{\be_j}$ are equal to $C$. We claim
that the $j$-fold left action of monodromy matrix elements results in
\begin{multline} \label{preleftactind}
     \<0| \Bigl[ \prod_{k = 1}^M C(\la_k) \Bigr]
        T^{\a_1}_{\be_1} (\la_{M+1}) \dots T^{\a_j}_{\be_j} (\la_{M+j})
	= \\
	  \sum_{\ell_1 = 1}^{M + \g_1}
	  \sum_{\substack{\ell_2 = 1 \\ \ell_2 \ne \ell_1}}^{M + \g_2}
	  \dots
	  \sum_{\substack{\ell_n = 1 \\ \ell_n \ne \ell_1, \dots,
	                  \ell_{n-1}}}^{M + \g_n}
	  \ph_1 (\ell_1, \g_1) \dots \ph_n (\ell_n, \g_n) \:
          \<0| \mspace{-9.0mu}
	       \prod_{\substack{k = 1 \\ k \ne \ell_1, \dots,
	                        \ell_n}}^{M + j}
	       \mspace{-18.0mu} C(\la_k) \epc
\end{multline}
where either
\begin{subequations}
\begin{align} \label{phina}
     \ph_n (\ell_n, \g_n) & =
        a(\la_{\ell_n}) c(\la_{M + \g_n}, \la_{\ell_n})
        \prod_{\substack{k = 1 \\ k \ne \ell_1, \dots,
	                 \ell_n}}^{M + \g_n}
           \frac{1}{b(\la_k, \la_{\ell_n})} \\[-4ex]
\intertext{or} \label{phind}
     \ph_n (\ell_n, \g_n) & =
        d(\la_{\ell_n}) c(\la_{\ell_n}, \la_{M + \g_n})
	\prod_{\substack{k = 1 \\ k \ne \ell_1, \dots,
	                 \ell_n}}^{M + \g_n}
           \frac{1}{b(\la_{\ell_n}, \la_k)} \epp
\end{align}
\end{subequations}%
\renewcommand{\arraystretch}{1.4}
\tabcolsep3mm
\begin{table}
\begin{center}
\begin{tabular}{c||cccccccc}
$j$ & 1 & 2 & 3 & 4 & 5 & 6 & 7 & 8 \\ \hline \hline
$\a_j$ & $\ab$ & $\ab$ & $\auf$ & $\auf$ & $\ab$ & $\auf$ & $\ab$ &
$\ab$ \\
$\be_j$ & $\ab$ & $\auf$ & $\ab$ & $\ab$ & $\ab$ & $\auf$ & $\auf$ &
$\ab$ \\ \hline
$\a_j^+$ & 3 & 4 & 6 &&&&& \\
$\be_j^-$ & 1 & 3 & 4 & 5 & 8 &&& \\ \hline
$\aqq_j^+$ & 6 & 4 & 3 &&&&& \\
$\bqq_j^-$ &&&& 1 & 3 & 4 & 5 & 8 \\ \hline
$\g_j$ & 1 & 3 & 3 & 4 & 4 & 5 & 6 & 8 \\ \hline
$\g_j^+$ & 3 & 5 & 7 &&&&& \\ \hline
$\g_j^-$ & 1 & 2 & 4 & 6 & 8 &&&
\end{tabular}
\end{center}%
\caption{\label{tab:ex1} Example for the definition of the various
sequences needed in the formulation and in the proof of lemma
\ref{lemma:leftact}. The spin pattern corresponds to the product
$D(\x_1) C(\x_2) B(\x_3) B(\x_4) D(\x_5) A(\x_6) C(\x_7) D(\x_8)$ of
monodromy matrix elements, m~=~8, $|\a^+| = 3$, $|\be^-| = 5$.}%
\end{table}%

Here we defined a sequence $(\g_n)_{n=1}^m$ by arranging all $\a_j^+$
and $\be_k^-$ in non-decreasing order (see table \ref{tab:ex1} for an
example).

\begin{proof}
Let us prove our statement above by induction over $j$. First of all
(\ref{preleftactind}) is satisfied for $j = 1$ by (\ref{ccca})-%
(\ref{cccb}) and the trivial formula
\begin{equation}
     \<0| \Bigl[ \prod_{k=1}^M C(\la_k) \Bigr] C(\la_{M+1})
        = \<0| \prod_{k=1}^{M+1} C(\la_k) \epp
\end{equation}

Next, let us assume that (\ref{preleftactind}) is true for some $j$
between 1 and $m-1$, and let us act with $T^{\a_{j+1}}_{\be_{j+1}}
(\la_{M + j + 1})$ from the right. We have to distinguish four
possible cases $(\a_{j+1}, \be_{j+1}) = (2,1), (1,1), (2,2), (1,2)$.

The case $(2,1)$. Then $T^{\a_{j+1}}_{\be_{j+1}} (\la_{M + j + 1}) =
C (\la_{M + j + 1})$, and
\begin{multline}
     \<0| \Bigl[ \prod_{k = 1}^M C(\la_k) \Bigr]
        T^{\a_1}_{\be_1} (\la_{M+1})
	  \dots T^{\a_{j+1}}_{\be_{j+1}} (\la_{M+j+1}) = \\
	  \sum_{\ell_1 = 1}^{M + \g_1}
	  \sum_{\substack{\ell_2 = 1 \\ \ell_2 \ne \ell_1}}^{M + \g_2}
	  \dots
	  \sum_{\substack{\ell_n = 1 \\ \ell_n \ne \ell_1, \dots,
	                  \ell_{n-1}}}^{M + \g_n}
	  \ph_1 (\ell_1, \g_1) \dots \ph_n (\ell_n, \g_n) \:
          \<0| \mspace{-9.0mu}
	       \prod_{\substack{k = 1 \\ k \ne \ell_1, \dots,
	                        \ell_n}}^{M + j +1}
	       \mspace{-18.0mu} C(\la_k) \epp
\end{multline}
This is of the same form as (\ref{preleftactind}), and the number of
sums $n$ is in accordance with our rule, since $\a_{j+1} = \ab = 2$
and $\be_{j+1} = \auf = 1$.

The case $(1,1)$. Now $T^{\a_{j+1}}_{\be_{j+1}} (\la_{M + j + 1}) =
A (\la_{M + j + 1})$, the number of sums increases by one in accordance
with our rule, and $\g_{n+1} = j + 1$. Using (\ref{ccca}),
(\ref{preleftactind}) we obtain
\begin{multline}
     \<0| \Bigl[ \prod_{k = 1}^M C(\la_k) \Bigr]
        T^{\a_1}_{\be_1} (\la_{M+1})
	  \dots T^{\a_{j+1}}_{\be_{j+1}} (\la_{M+j+1}) = \\
	  \sum_{\ell_1 = 1}^{M + \g_1}
	  \sum_{\substack{\ell_2 = 1 \\ \ell_2 \ne \ell_1}}^{M + \g_2}
	  \dots
	  \sum_{\substack{\ell_n = 1 \\ \ell_n \ne \ell_1, \dots,
	                  \ell_{n-1}}}^{M + \g_n}
	  \sum_{\substack{\ell_{n+1} = 1 \\ \ell_{n+1} \ne \ell_1, \dots,
	                  \ell_n}}^{M + j + 1}
	  \ph_1 (\ell_1, \g_1) \dots \ph_n (\ell_n, \g_n) \\
          a(\la_{\ell_{n+1}}) c(\la_{M+j+1}, \la_{\ell_{n+1}})
	  \biggr[ \prod_{\substack{k = 1 \\
	                  k \ne \ell_1, \dots, \ell_{n+1}}}^{M+j+1}
	         \frac{1}{b(\la_k, \la_{\ell_{n+1}})} \biggl]
          \<0| \mspace{-9.0mu}
	       \prod_{\substack{k = 1 \\ k \ne \ell_1, \dots,
	                        \ell_{n+1}}}^{M + j + 1}
	       \mspace{-18.0mu} C(\la_k) = \\
	  \sum_{\ell_1 = 1}^{M + \g_1}
	  \sum_{\substack{\ell_2 = 1 \\ \ell_2 \ne \ell_1}}^{M + \g_2}
	  \dots
	  \sum_{\substack{\ell_{n+1} = 1 \\ \ell_{n+1} \ne \ell_1, \dots,
	                  \ell_n}}^{M + \g_{n+1}}
	  \ph_1 (\ell_1, \g_1) \dots \ph_{n+1} (\ell_{n+1}, \g_{n+1})
          \<0| \mspace{-9.0mu}
	       \prod_{\substack{k = 1 \\ k \ne \ell_1, \dots,
	                        \ell_{n+1}}}^{M + j + 1}
	       \mspace{-18.0mu} C(\la_k)
\end{multline}
which is again of the form (\ref{preleftactind}). We substituted
$\g_{n+1} = j + 1$ and used (\ref{phina}) in the last equation. The cases
$(2,2)$ and $(1,2)$, being very similar, are left to the reader.
\end{proof}

Next we set $j = m$ in equation (\ref{preleftactind}). Because of the
constraint (\ref{constraintab}), we have $m$ sums on the right hand side
and thus,
\begin{multline} \label{preleftact}
     \<0| \Bigl[ \prod_{k = 1}^M C(\la_k) \Bigr]
        T^{\a_1}_{\be_1} (\la_{M+1}) \dots T^{\a_m}_{\be_m} (\la_{M+m})
	= \\
	  \sum_{\ell_1 = 1}^{M + \g_1}
	  \sum_{\substack{\ell_2 = 1 \\ \ell_2 \ne \ell_1}}^{M + \g_2}
	  \dots
	  \sum_{\substack{\ell_m = 1 \\ \ell_m \ne \ell_1, \dots,
	                  \ell_{m-1}}}^{M + \g_m}
	  \ph_1 (\ell_1, \g_1) \dots \ph_m (\ell_m, \g_m) \:
          \<0| \mspace{-9.0mu}
	       \prod_{\substack{k = 1 \\ k \ne \ell_1, \dots,
	                        \ell_m}}^{M + m}
	       \mspace{-18.0mu} C(\la_k) \epp
\end{multline}
The right hand side of this equation becomes explicit after deciding
for each function $\ph_j$ if (\ref{phina}) or (\ref{phind}) applies.
In order to do so we note that a unique partition $(\{\g^+\},\{\g^-\})
\in p_2 ({\mathbb Z}_m)$ exists such that
\begin{equation}
       \g_{\g_j^+} = \a_j^+ \epc \qqd \g_{\g_j^-} = \be_j^- \epc \qqd
       \g_j^- < \g_k^+\ \text{if}\ \be_j^- = \a_k^+ \epp
\end{equation}
For an example see again table \ref{tab:ex1}. Now each factor $\ph_j
(\ell_j, \g_j)$ with $\g_j \ne \g_k$ for $j \ne k$ stems either from
the application of $A$ and must be replaced by (\ref{phina}) or
it stems from the application of $D$ and must be replaced by
(\ref{phind}). In the former case $j \in \{\g^+\}$ while in the latter
$j \in \{\g^-\}$. For a pair of equal $\g$s, say $\g_j = \g_{j+1}$,
which stems from the application of an operator $B$, $\ph_j
(\ell_j, \g_j)$ must be replaced by (\ref{phind}) and $\ph_{j+1}
(\ell_{j+1}, \g_{j+1})$ is determined by (\ref{phina}) (see
(\ref{cccb})). Hence, $j \in \{\g^-\}$ and $j + 1 \in \{\g^+\}$. Taking
the three cases together, we see that for every $j \in \{\g^+\}$ we have
to substitute $\ph_j$ using (\ref{phina}), while for every $j \in
\{\g^-\}$ equation (\ref{phind}) applies. Thus, we have derived the
following
\begin{lemma} \cite{GKS05} The general left action formula, version 1,
\begin{multline} \label{genact}
     \<0| \Bigl[ \prod_{k = 1}^M C(\la_k) \Bigr]
        T^{\a_1}_{\be_1} (\la_{M+1}) \dots T^{\a_m}_{\be_m} (\la_{M+m})
	= \\
	  \sum_{\ell_1 = 1}^{M + \g_1}
	  \sum_{\substack{\ell_2 = 1 \\ \ell_2 \ne \ell_1}}^{M + \g_2}
	  \dots
	  \sum_{\substack{\ell_m = 1 \\ \ell_m \ne \ell_1, \dots,
	                  \ell_{m-1}}}^{M + \g_m}
	  \biggl[ \prod_{j=1}^{|\a^+|} a(\la_{\ell_{\g_j^+}})
	          c(\la_{M + \a_j^+}, \la_{\ell_{\g_j^+}})
	          \prod_{\substack{k = 1 \\ k \ne \ell_1, \dots,
	 	                   \ell_{\g_j^+}}}^{M + \a_j^+}
		  \frac{1}{b(\la_k, \la_{\ell_{\g_j^+}})} \biggr] \\
	  \biggl[ \prod_{j=1}^{m-|\a^+|} d(\la_{\ell_{\g_j^-}})
	          c(\la_{\ell_{\g_j^-}}, \la_{M + \be_j^-})
	          \prod_{\substack{k = 1 \\ k \ne \ell_1, \dots,
		                   \ell_{\g_j^-}}}^{M + \be_j^-}
		  \frac{1}{b(\la_{\ell_{\g_j^-}}, \la_k)} \biggr]
          \<0| \mspace{-9.0mu}
	       \prod_{\substack{k = 1 \\ k \ne \ell_1, \dots,
	                        \ell_m}}^{M + m}
	       \mspace{-18.0mu} C(\la_k) \epp
\end{multline}
\end{lemma}

Let us briefly rest with this formula. Equation (\ref{genact}) is
a very compact formula for the general left action of a ($S^z$-%
conserving) string of monodromy matrix elements on a state of the form
(\ref{cform}). In a sense (\ref{genact}) seems even more generic than
the formula (\ref{gle}) that we are aiming at, since the coefficient
under the sum in (\ref{genact}) is entirely defined in terms of the
functions $b(\la, \m)$ and $c(\la, \m)$ that determine the $R$-matrix
and the vacuum expectation values $a(\la)$ and $d(\la)$ of the (quantum)
transfer matrix. On the other hand we face a rather complicated
dependence of the limits of summation on the matrix indices of the
density matrix elements and the explicit appearance of the partition
$(\{\g^+\},\{\g^-\})$ in the coefficient under the sum.

Starting from (\ref{genact}) we can now derive (\ref{gle}) by rather
elementary manipulations. Let us nevertheless sketch the main steps.
The notation
\begin{subequations}
\label{lpmmp}
\begin{align} \label{lpp}
     |\ell^{++}| & = \card \{1, \dots, M\} \cap
                     \{\ell_{\g_j^+}\}_{j=1}^{|\a^+|} \epc \\ \label{lmp}
     |\ell^{-+}| & = \card \{M + 1, \dots, M + m\} \cap
                     \{\ell_{\g_j^+}\}_{j=1}^{|\a^+|} \epc \\ \label{lpm}
     |\ell^{+-}| & = \card \{1, \dots, M\} \cap
                     \{\ell_{\g_j^-}\}_{j=1}^{m - |\a^+|} \epc \\
     |\ell^{--}| & = \card \{M + 1, \dots, M + m\} \cap
                     \{\ell_{\g_j^-}\}_{j=1}^{m - |\a^+|} \label{lmm}
\end{align}
\end{subequations}
turns out to be convenient in the calculation. We shall denote the
coefficient under the sum on the right hand side of (\ref{genact}) as
$S = S_{\ell_1, \dots, \ell_m}$. Then
\begin{align} \label{defs}
     S  = & \biggl[ \prod_{j=1}^{|\a^+|} a(\la_{\ell_{\g_j^+}})
	            c(\la_{M + \a_j^+}, \la_{\ell_{\g_j^+}})
	            \prod_{\substack{k = 1 \\ k \ne \ell_1, \dots,
		           \ell_{\g_j^+}}}^{M + \a_j^+}
		    \frac{1}{b(\la_k, \la_{\ell_{\g_j^+}})} \biggr]
		    \notag \\
	  & \biggl[ \prod_{j=1}^{m-|\a^+|} d(\la_{\ell_{\g_j^-}})
	            c(\la_{\ell_{\g_j^-}}, \la_{M + \be_j^-})
	            \prod_{\substack{k = 1 \\ k \ne \ell_1, \dots,
		                     \ell_{\g_j^-}}}^{M + \be_j^-}
		    \frac{1}{b(\la_{\ell_{\g_j^-}}, \la_k)} \biggr]
		    \notag \\
        = & (-1)^{|\ell^{+-}|}
	      \biggl[ \prod_{j = 1}^{m - |\a^+|}
	              \fb_M (\la_{\ell_{\g_j^-}}) \biggr]
	      \biggl[ \prod_{j = 1}^m a (\la_{\ell_j})
	              \prod_{\substack{k = 1\\ k \ne \ell_j}}^M
		      \frac{1}{b(\la_k, \la_{\ell_j})} \biggr] U V \epc
\end{align}
where
\begin{subequations}
\label{defuv}
\begin{align}
     U = & \biggl[ \prod_{j=1}^{|\a^+|} \prod_{k = 1}^{\g_j^+ - 1}
		   b(\la_{\ell_k}, \la_{\ell_{\g_j^+}}) \biggr]
	   \biggl[ \prod_{j=1}^{m-|\a^+|} \prod_{k = 1}^{\g_j^- - 1}
		   b(\la_{\ell_{\g_j^-}}, \la_{\ell_k}) \biggr] \epc
		   \\[2ex]
     V = & \biggl[ \prod_{j=1}^{|\a^+|}
	            c(\la_{M + \a_j^+}, \la_{\ell_{\g_j^+}})
	            \prod_{\substack{k = 1 \\
		           k + M \ne \ell_{\g_j^+}}}^{\a_j^+}
		    \frac{1}{b(\x_k, \la_{\ell_{\g_j^+}})} \biggr]
		    \notag \\[-2ex] & \mspace{90.mu}
	   \biggl[ \prod_{j=1}^{m-|\a^+|}
	            c(\la_{\ell_{\g_j^-}}, \la_{M + \be_j^-})
	            \prod_{\substack{k = 1 \\
		           k + M \ne \ell_{\g_j^-}}}^{\be_j^-}
		    \frac{1}{b(\la_{\ell_{\g_j^-}}, \x_k)} \biggr] \epp
\end{align}
\end{subequations}
In the second equation (\ref{defs}) we have pulled out several factors
and we have introduced the notation
\begin{equation} \label{defam}
     \fb_M (\la) = \frac{d(\la)}{a(\la)} \prod_{k = 1}^M
                      \frac{\sh(\la - \la_k + \h)}{\sh(\la - \la_k - \h)}
		      \epp
\end{equation}
Notice that we did not assume the $\la_j$ to satisfy any equation here.
In the next step we simplify the terms $U$ and $V$ defined in 
(\ref{defuv}).

Simplification of $U$.
\begin{align} \label{fineu}
     U = & \biggl[ \prod_{j=1}^{|\a^+|} \prod_{k = 1}^{j - 1}
		   b(\la_{\ell_{\g_k^+}}, \la_{\ell_{\g_j^+}})
	   \mspace{-2.mu}
		   \prod_{k = 1}^{\g_j^+ - j}
		   b(\la_{\ell_{\g_k^-}}, \la_{\ell_{\g_j^+}}) \biggr]
	   \mspace{-5.mu}
	   \biggl[ \prod_{j=1}^{m-|\a^+|} \prod_{k = 1}^{j - 1}
		   b(\la_{\ell_{\g_j^-}}, \la_{\ell_{\g_k^-}})
	   \mspace{-2.mu}
		   \prod_{k = 1}^{\g_j^- - j}
		   b(\la_{\ell_{\g_j^-}}, \la_{\ell_{\g_k^+}}) \biggr]
		   \notag \\[2ex]
       = & \biggl[ \prod_{j=1}^{m-|\a^+|} \prod_{k = 1}^{j - 1}
		   b(\la_{\ell_{\g_j^-}}, \la_{\ell_{\g_k^-}}) \biggr]
           \biggl[ \prod_{j=1}^{m-|\a^+|} \prod_{k=1}^{|\a^+|}
		   b(\la_{\ell_{\g_j^-}}, \la_{\ell_{\g_k^+}}) \biggr]
           \biggl[ \prod_{j=1}^{|\a^+|} \prod_{k = 1}^{j - 1}
		   b(\la_{\ell_{\g_k^+}}, \la_{\ell_{\g_j^+}}) \biggr]
		   \epp
\end{align}
Here we used in the first equation that $\g_j^\pm - j$ is the number
of $\g^\mp$s smaller than $\g_j^\pm$. In the second equation we
combined the second and fourth factor on the left hand side into
the second factor on the right hand side. Now we have removed the
dependence on $\g_j^\pm$ from the upper limits of the products.

Simplification of $V$. We first observe that
\begin{equation} \label{vextractden}
     V \prod_{j=1}^m \prod_{\substack{k = 1\\ k + M \ne \ell_j}}^m
	\sh(\la_{\ell_j} - \x_k) = \sh^m (\h)
	\biggl[ \prod_{j=1}^{|\a^+|} V_j^{(+)} \biggr]
	\biggl[ \prod_{j=1}^{m - |\a^+|} V_j^{(-)} \biggr]
\end{equation}
with
\begin{subequations}
\begin{align}
     V_j^{(+)} = & \frac{-1}{\sh(\la_{\ell_{\g_j^+}} - \x_{\a_j^+} - \h)}
	           \prod_{\substack{k = 1 \\
		          k + M \ne \ell_{\g_j^+}}}^{\a_j^+}
			  \mspace{-10.mu}
                   \frac{\sh(\la_{\ell_{\g_j^+}} - \x_k - \h)}
                        {\sh(\la_{\ell_{\g_j^+}} - \x_k)}
			  \mspace{-10.mu}
	           \prod_{\substack{k = 1 \\
		          k + M \ne \ell_{\g_j^+}}}^m
			  \mspace{-10.mu}
                          \sh(\la_{\ell_{\g_j^+}} - \x_k) \epc \\
     V_j^{(-)} = & \frac{1}{\sh(\la_{\ell_{\g_j^-}} - \x_{\be_j^-} + \h)}
	           \prod_{\substack{k = 1 \\
		          k + M \ne \ell_{\g_j^-}}}^{\be_j^-}
			  \mspace{-10.mu}
                   \frac{\sh(\la_{\ell_{\g_j^-}} - \x_k + \h)}
                        {\sh(\la_{\ell_{\g_j^-}} - \x_k)}
			  \mspace{-10.mu}
	           \prod_{\substack{k = 1 \\
		          k + M \ne \ell_{\g_j^-}}}^m
			  \mspace{-10.mu}
                          \sh(\la_{\ell_{\g_j^-}} - \x_k) \epp
\end{align}
\end{subequations}
Now we have to distinguish different cases.

(i) $\ell_{\g_j^+} \in \{1, \dots, M\}\ \then\ k + M \ne \ell_{\g_j^+}$,
and
\begin{equation}
     V_j^{(+)} = - \prod_{k = 1}^{\a_j^+ - 1}
                   \sh(\la_{\ell_{\g_j^+}} - \x_k - \h)
	           \prod_{k = \a_j^+ + 1}^m
		   \sh(\la_{\ell_{\g_j^+}} - \x_k) \epp
\end{equation}

(ii) $\ell_{\g_j^+} \in \{M + 1, \dots, M + \g_{\g_j^+} = M + \a_j^+\}\
\then\ k + M \ne \ell_{\g_j^+}\ \text{for all}\ k > \a_j^+$, and
\begin{equation}
     V_j^{(+)} = \frac{1}{\sh(\h)}
                 \prod_{k = 1}^{\a_j^+ - 1}
                 \sh(\la_{\ell_{\g_j^+}} - \x_k - \h)
	         \prod_{k = \a_j^+ + 1}^m
		 \sh(\la_{\ell_{\g_j^+}} - \x_k) \epp
\end{equation}

(iii) $\ell_{\g_j^-} \in \{1, \dots, M\}\ \then\ k + M \ne
\ell_{\g_j^-}$, and
\begin{equation}
     V_j^{(-)} = \prod_{k = 1}^{\be_j^- - 1}
                 \sh(\la_{\ell_{\g_j^-}} - \x_k + \h)
	         \prod_{k = \be_j^- + 1}^m
		 \sh(\la_{\ell_{\g_j^-}} - \x_k) \epp
\end{equation}

(iv) $\ell_{\g_j^-} \in \{M + 1, \dots, M + \g_{\g_j^-} = M + \be_j^-\}\
\then\ k + M \ne \ell_{\g_j^-}\ \text{for all}\ k > \be_j^-$, and
\begin{equation} \label{vmiv}
     V_j^{(-)} = \frac{1}{\sh(\h)}
                 \prod_{k = 1}^{\be_j^- - 1}
                 \sh(\la_{\ell_{\g_j^-}} - \x_k + \h)
	         \prod_{k = \be_j^- + 1}^m
		 \sh(\la_{\ell_{\g_j^-}} - \x_k) \epp
\end{equation}

Combining (\ref{vextractden})-(\ref{vmiv}) and using the notation
(\ref{lpmmp}) we conclude that
\begin{multline} \label{finev}
     V = (-1)^{|\ell^{++}|}
         \bigl( \sh(\h) \bigr)^{|\ell^{++}| + |\ell^{+-}|}
	 \biggl[ \prod_{j=1}^m
	         \prod_{\substack{k = 1\\ k + M \ne \ell_j}}^m
	         \frac{1}{\sh(\la_{\ell_j} - \x_k)} \biggr] \\
	 \biggl[ \prod_{j=1}^{|\a^+|}
                 \prod_{k = 1}^{\a_j^+ - 1}
                 \sh(\la_{\ell_{\g_j^+}} - \x_k - \h)
	         \prod_{k = \a_j^+ + 1}^m
		 \sh(\la_{\ell_{\g_j^+}} - \x_k) \biggr] \\
	 \biggl[ \prod_{j=1}^{m - |\a^+|}
                 \prod_{k = 1}^{\be_j^- - 1}
                 \sh(\la_{\ell_{\g_j^-}} - \x_k + \h)
	         \prod_{k = \be_j^- + 1}^m
		 \sh(\la_{\ell_{\g_j^-}} - \x_k) \biggr] \epc
\end{multline}
which is again of a form where the range of the products does not
depend on $\g_j^\pm$ anymore.

Let us now return back to the sum on the right hand side of
(\ref{genact}). The coefficient under the sum is our $S$, equation
(\ref{defs}), where $U$ and $V$ are given by (\ref{fineu}) and
(\ref{finev}). The summation variables $\ell_j$ range from 1 to $M +
\g_j$, i.e.,
\begin{subequations}
\begin{align}
     & \ell_{\g_j^+} \in \{1, \dots, M + \a_j^+\} \epc \qqd
        j = 1, \dots, |\a^+| \epc \\
     & \ell_{\g_j^-} \in \{1, \dots, M + \be_j^-\} \epc \qqd
        j = 1, \dots, m - |\a^+| \epp
\end{align}
\end{subequations}
Let us now assume that $\ell_{\g_j^+} \in \{M + \a_j^+  + 1, \dots,
M + m\}$ for some $j$. Then $\la_{\ell_{\g_j^+}} = \x_{\ell_{\g_j^+ - M}}
\in \{\x_{\a_j^+ + 1}, \dots, \x_m\}$, and it follows from
(\ref{finev}) that $S = 0$. Similarly, $S = 0$ if $\ell_{\g_j^-} \in
\{M + \be_j^-  + 1, \dots, M + m\}$ for some $j$. Hence, we may extend
the upper limit of summation to $M + m$ in all sums on the right hand
side of (\ref{genact}), since this just means to add zeros to the sum.

Then the sum is symmetric in the summation variables and may be written
as
\begin{equation} \label{sumsym}
        \sum_{\substack{\ell_1, \dots, \ell_m = 1\\
	      \ell_j \ne \ell_k\ \text{for}\ j \ne k}}^{M+m}
	      S_{\ell_1, \dots, \ell_m} \qd
          \<0| \mspace{-9.0mu}
	       \prod_{\substack{k = 1 \\ k \ne \ell_1, \dots,
	                        \ell_m}}^{M + m}
	       \mspace{-18.0mu} C(\la_k) \epp
\end{equation}
Since the product over $C$s is symmetric in the $\ell_j$ as well, we
may now permute the summation variables in S according to our
convenience. Substituting
\begin{subequations}
\label{substsumvar}
\begin{align}
     \ell_{\g_j^+}\ & \longrightarrow\ \ell_{|\a^+| - j + 1} \epc &
        j & = 1, \dots, |\a^+| \epc \\
     \ell_{\g_j^-}\ & \longrightarrow\ \ell_{|\a^+| + j} \epc &
        j & = 1, \dots, m - |\a^+| \epc
\end{align}
\end{subequations}
we can simplify $U$,
\begin{multline}
     U = \biggl[ \prod_{j=1}^{m-|\a^+|} \prod_{k = 1}^{j - 1}
                 b(\la_{\ell_{|\a^+| + j}},
		   \la_{\ell_{|\a^+| + k}}) \biggr]
         \biggl[ \prod_{j=1}^{m-|\a^+|} \prod_{k=1}^{|\a^+|}
	         b(\la_{\ell_{|\a^+| + j}},
		   \la_{\ell_{|\a^+| - k + 1}}) \biggr] \\
         \biggl[ \prod_{j=1}^{|\a^+|} \prod_{k = 1}^{j - 1}
	         b(\la_{\ell_{|\a^+| - k + 1}},
		   \la_{\ell_{|\a^+| - j + 1}}) \biggr]
       = \prod_{1 \le j < k \le m} b(\la_{\ell_k}, \la_{\ell_j}) \epc
\end{multline}
and $S$ becomes
\begin{multline} \label{fines}
     S = \bigl(- \sh(\h) \bigr)^{|\ell^{++}| + |\ell^{+-}|}
	 \biggl[ \prod_{j = |\a^+| + 1}^m
	         \fb_M (\la_{\ell_j}) \biggr]
	 \biggl[ \prod_{j = 1}^m a (\la_{\ell_j})
	         \prod_{\substack{k = 1\\ k \ne \ell_j}}^M
		 \frac{1}{b(\la_k, \la_{\ell_j})} \biggr] \\
         \biggl[ \prod_{1 \le j < k \le m}
	         b(\la_{\ell_k}, \la_{\ell_j}) \biggr]
	 \biggl[ \prod_{j=1}^m
	         \prod_{\substack{k = 1\\ k + M \ne \ell_j}}^m
	         \frac{1}{\sh(\la_{\ell_j} - \x_k)} \biggr] \\
	 \biggl[ \prod_{j=1}^{|\a^+|}
                 \prod_{k = 1}^{\a_j^+ - 1}
                 \sh(\la_{\ell_{|\a^+| - j + 1}} - \x_k - \h)
	         \prod_{k = \a_j^+ + 1}^m
		 \sh(\la_{\ell_{|\a^+| - j + 1}} - \x_k) \biggr] \\
	 \biggl[ \prod_{j=1}^{m - |\a^+|}
                 \prod_{k = 1}^{\be_j^- - 1}
                 \sh(\la_{\ell_{|\a^+| + j}} - \x_k + \h)
	         \prod_{k = \be_j^- + 1}^m
		 \sh(\la_{\ell_{|\a^+| + j}} - \x_k) \biggr] \epp
\end{multline}
Here we introduce
\begin{subequations}
\begin{align}
     \aqq_j^+ & = \a^+_{|\a^+| - j + 1} \epc &
                  j & = 1, \dots, |\a^+| \epc \\
     \bqq_j^- & = \be^-_{j - |\a^+|} \epc &
                  j & = |\a^+| + 1, \dots, m
\end{align}
\end{subequations}
and shift the indices of the products in the last two terms in
(\ref{fines}). Setting $|\ell^+| = |\ell^{++}| + |\ell^{+-}|$ and
substituting $S$ back into (\ref{sumsym}) we obtain the desired result
(\ref{gle}).

Notice that, because of (\ref{lpp}) and (\ref{lpm}), we have
\begin{equation}
     |\ell^+| = \card \{1, \dots, M\} \cap \{\ell_j\}_{j=1}^m
\end{equation}
in the new summation  variables (\ref{substsumvar}).

\clearpage
\Appendix{The coefficients in the generating function}
\label{app:emptgen} \noindent
In this appendix we give a proof of theorem \ref{theo:coefff}. For fixed
$n \in \{0, 1, \dots, m\}$ consider the $n$th term
$D^{\a_1 \dots \a_m}_{\a_1 \dots \a_m} (\x_1, \dots, \x_m)$ under the
sum (\ref{fsumd}). It satisfies $\a_1 + \dots + \a_m = 2m - n$, which
means it has $|\a^+| = n$ up-spins. Since the sequences of upper and
lower indices of $D$ are identical, it follows that $\{\aqq^+_j\}_{j=1}^n
\cap \{\bqq^-_j\}_{j=n+1}^m = \emptyset$. Hence, $(\{\aqq^+\},
\{\bqq^-\}) \in p_2 ({\mathbb Z}_m)$, and a unique $P \in
{\mathfrak S}^m$ exists, such that
\begin{equation} \label{paqqbqq}
     P \aqq_j^+ = j \epc \qd j = 1, \dots, n \epc \qqd
     P \bqq_j^- = j \epc \qd j = n + 1, \dots, m \epp
\end{equation}
Let us define a sequence $(\e_j)_{j=1}^m$ by
\begin{equation}
     \e_j = \begin{cases}
            - \2 & j = 1, \dots, n \\
	    + \2 & j = n + 1, \dots, m \epp
	    \end{cases}
\end{equation}
Then
\begin{multline} \label{ffepsp}
     \biggl[ \prod_{j=1}^n
	     \prod_{k=1}^{\aqq_j^+ - 1} \sh(\om_j - \x_k - \h)
	     \prod_{k = \aqq_j^+ + 1}^m \sh(\om_j - \x_k) \biggr]
	     \\[-1ex]
     \biggl[ \prod_{j = n + 1}^{m}
	     \prod_{k=1}^{\bqq_j^- - 1} \sh(\om_j - \x_k + \h)
	     \prod_{k = \bqq_j^- + 1}^m \sh(\om_j - \x_k) \biggr] \\[1ex]
     = \prod_{1 \le j < k \le m}
       \sh(\om_{Pk} - \x_j + 2 \e_{Pk} \h) \sh(\om_{Pj} - \x_k)
\end{multline}
which considerably simplifies the integral formula for
$D^{\a_1 \dots \a_m}_{\a_1 \dots \a_m} (\x_1, \dots, \x_m)$.

We wish to express the product $\prod_{1 \le j < k \le m}
\sh(\om_k - \om_j + \h)$ as well in terms of $\om_{Pj}$. For this
purpose we note that the range ${\cal M}$ of the multiplication
variables consists of all ordered pairs $(j,k)$ and is, in general, not
invariant under P. However, if $(\ell, n) \in P {\cal M}$, then
$(n, \ell) \notin P {\cal M}$. It follows that
\begin{multline} \label{prodepsp}
     \prod_{1 \le j < k \le m} \sh(\om_k - \om_j + \h)
        = \prod_{1 \le j < k \le m}
	  \sh \bigl(\sign(Pk - Pj)(\om_{Pk} - \om_{Pj}) + \h \bigr) \\
        = \biggl[ \prod_{1 \le j < k \le m}
	  \frac{\om_j - \om_k}{\om_{Pj} - \om_{Pk}} \biggr]
	  \biggl[ \prod_{1 \le j < k \le m}
	  \sh \bigl( \om_{Pk} - \om_{Pj} + \sign(Pk - Pj) \h \bigr)
	  \biggr] \\[1ex]
        = \sign(P) \prod_{1 \le j < k \le m}
	  \sh \bigl( \om_{Pk} - \om_{Pj} + 2 \e_{Pk} \h \bigr) \epp
\end{multline}
Here we used that the first product in the second line is a ratio of two
van der Monde determinants. We further employed the identity
\begin{equation}
     \sign(Pk - Pj) = 2 \e_{Pk} \epc \qd \text{for $j < k$,}
\end{equation}
which can be inferred from the definitions of $P$ and $(\e_j)$.
Inserting (\ref{ffepsp}) and (\ref{prodepsp}) into (\ref{densint}) and
performing the Trotter limit we obtain
\begin{align} \label{densintaa}
     D^{\a_1 \dots \a_m}_{\a_1 \dots \a_m} & (\x_1, \dots, \x_m) =
        \notag \\ & \mspace{-18.mu}
	\biggl[ \prod_{j=1}^n \int_{\cal C}
	    \frac{d \om_j}{2 \p \i (1 + \fa (\om_j))} \biggr]
            \biggl[ \prod_{j = n + 1}^{m} \int_{\cal C}
	    \frac{d \om_j}{2 \p \i (1 + \faq (\om_j))} \biggr]
            \frac{\det( - G(\om_j, \x_k))}
	     {\prod_{1 \le j < k \le m} \sh(\x_j - \x_k)}
        \notag \\ & \mspace{-18.mu}
        \sign(P) \prod_{1 \le j < k \le m}
	\frac{\sh(\om_{Pk} - \x_j + 2 \e_{Pk} \h) \sh(\om_{Pj} - \x_k)}
	     {\sh \bigl( \om_{Pk} - \om_{Pj} + 2 \e_{Pk} \h \bigr)} \epp
\end{align}
This is a nice alternative form of the integral formula for the density
matrix elements in the special case of only diagonal local operators 
$e^1_1$ and $e^2_2$.

The reason why we introduced the seemingly more complicated expression
(\ref{prodepsp}) into (\ref{densintaa}) is that in this form the
invariance properties with respect to certain variation of the
permutation $P$ become obvious. Namely, if $Q \in {\mathfrak S}^m$ with
$Q \{1, \dots, n\} = \{1, \dots, n\}$, then
\begin{align} \label{densintaainv}
     D^{\a_1 \dots \a_m}_{\a_1 \dots \a_m} & (\x_1, \dots, \x_m) =
        \notag \\ & \mspace{-18.mu}
	\biggl[ \prod_{j=1}^n \int_{\cal C}
	    \frac{d \om_j}{2 \p \i (1 + \fa (\om_j))} \biggr]
            \biggl[ \prod_{j = n + 1}^{m} \int_{\cal C}
	    \frac{d \om_j}{2 \p \i (1 + \faq (\om_j))} \biggr]
            \frac{\det( - G(\om_j, \x_k))}
	     {\prod_{1 \le j < k \le m} \sh(\x_j - \x_k)}
        \notag \\ & \mspace{-18.mu}
        \sign(QP) \prod_{1 \le j < k \le m}
	\frac{\sh(\om_{QPk} - \x_j + 2 \e_{QPk} \h)
	      \sh(\om_{QPj} - \x_k)}
	     {\sh \bigl( \om_{QPk} - \om_{QPj} + 2 \e_{QPk} \h \bigr)}
	      \epc
\end{align}
where we have used that, by definition of the permutation $Q$, the
identity $\e_{QPk} = \e_{Pk}$ holds for all $k \in \{1, \dots, m\}$.

Next we would like to carry out the summation in (\ref{fsumd}). The
sum can be equivalently written as a sum over all partitions
$(\{\aqq^+\}, \{\bqq^-\}) \in p_2 ({\mathbb Z}_m)$ with $|\a^+| = n$.
As is clear from (\ref{paqqbqq}) every such partition uniquely
corresponds to a permutation $P \in {\mathfrak S}^m$ which hence will
be denoted $P_{\{\aqq^+\}}$. It follows that
\begin{multline} \label{fsumalt1}
     f_n (\x_1, \dots, \x_m) = \\[1ex]
	\biggl[ \prod_{j=1}^n \int_{\cal C}
	    \frac{d \om_j}{2 \p \i (1 + \fa (\om_j))} \biggr]
            \biggl[ \prod_{j = n + 1}^{m} \int_{\cal C}
	    \frac{d \om_j}{2 \p \i (1 + \faq (\om_j))} \biggr]
            \frac{\det( - G(\om_j, \x_k))}
	     {\prod_{1 \le j < k \le m} \sh(\x_j - \x_k)} \\
     \sum_{\substack{(\{\aqq^+\}, \{\bqq^-\}) \in p_2 ({\mathbb Z}_m) \\
                     |\a^+| = n}}
		     \mspace{-45.mu}
		     \sign(QP_{\{\aqq^+\}})
		     \mspace{-9.mu}
        \prod_{1 \le j < k \le m}
	\frac{\sh(\om_{QP_{\{\aqq^+\}}k} - \x_j
	          + 2 \e_{QP_{\{\aqq^+\}}k} \h)
	      \sh(\om_{QP_{\{\aqq^+\}}j} - \x_k)}
	     {\sh \bigl( \om_{QP_{\{\aqq^+\}}k} - \om_{QP_{\{\aqq^+\}}j}
	                 + 2 \e_{QP_{\{\aqq^+\}}k} \h \bigr)}
\end{multline}
for all $Q \in {\mathfrak S}^m$ which leave the set $\{1, \dots, n\}$
invariant. These $Q$ form a subgroup, say ${\mathfrak U} \subset
{\mathfrak S}^m$. If now $\{\aqq^+\} \ne \{\aqq^{+ \prime}\}$ but
$|\a^+| = |\a^{+ \prime} | = n$, then $P_{\{\aqq^+\}} \ne
P_{\{\aqq^{+ \prime}\}}$, and $QP_{\{\aqq^+\}} \{\aqq^+\} =
\{1, \dots, n\}$ but $QP_{\{\aqq^{+ \prime}\}} \{\aqq^+\} \ne
\{1, \dots, n\}$, which means that ${\mathfrak U} P_{\{\aqq^+\}} \cap
{\mathfrak U} P_{\{\aqq^{+ \prime}\}} = \emptyset$. Thus, ${\mathfrak U}$
determines a decomposition of the symmetric group ${\mathfrak S}^m$ into
cosets,
\begin{equation}
     {\mathfrak S}^m =
        \bigcup_{\substack{(\{\aqq^+\}, \{\bqq^-\})
	                   \in p_2 ({\mathbb Z}_m) \\ |\a^+| = n}}
        {\mathfrak U} P_{\{\aqq^+\}} \epp
\end{equation}

With this information we can average equation (\ref{fsumalt1}) over
the subgroup ${\mathfrak U}$ which has $n! (m - n)!$ elements,
\begin{multline} \label{fsumalt2}
     f_n (\x_1, \dots, \x_m) = 
	\biggl[ \prod_{j=1}^n \int_{\cal C}
	    \frac{d \om_j}{2 \p \i (1 + \fa (\om_j))} \biggr]
            \biggl[ \prod_{j = n + 1}^{m} \int_{\cal C}
	    \frac{d \om_j}{2 \p \i (1 + \faq (\om_j))} \biggr] \\
            \frac{\det( - G(\om_j, \x_k))}
	     {\prod_{1 \le j < k \le m} \sh(\x_j - \x_k)} \:
     \frac{1}{n! (m - n)!} \: \sum_{Q \in {\mathfrak U}} \:
     \sum_{\substack{(\{\aqq^+\}, \{\bqq^-\}) \in p_2 ({\mathbb Z}_m) \\
                     |\a^+| = n}}
		     \sign(QP_{\{\aqq^+\}}) \\
        \prod_{1 \le j < k \le m}
	\frac{\sh(\om_{QP_{\{\aqq^+\}}k} - \x_j
	          + 2 \e_{QP_{\{\aqq^+\}}k} \h)
	      \sh(\om_{QP_{\{\aqq^+\}}j} - \x_k)}
	     {\sh \bigl( \om_{QP_{\{\aqq^+\}}k} - \om_{QP_{\{\aqq^+\}}j}
	                 + 2 \e_{QP_{\{\aqq^+\}}k} \h \bigr)}
			 \mspace{90.mu} \\[3ex] =
	\biggl[ \prod_{j=1}^n \int_{\cal C}
	    \frac{d \om_j}{2 \p \i (1 + \fa (\om_j))} \biggr]
            \biggl[ \prod_{j = n + 1}^{m} \int_{\cal C}
	    \frac{d \om_j}{2 \p \i (1 + \faq (\om_j))} \biggr]
            \frac{\det( - G(\om_j, \x_k))}
	     {\prod_{1 \le j < k \le m} \sh(\x_j - \x_k)} \\
     \frac{1}{n! (m - n)!} \sum_{P \in {\mathfrak S}^m} \sign(P)
        \prod_{1 \le j < k \le m}
	\frac{\sh(\om_{Pk} - \x_j + 2 \e_{Pk} \h) \sh(\om_{Pj} - \x_k)}
	     {\sh \bigl( \om_{Pk} - \om_{Pj} + 2 \e_{Pk} \h \bigr)} \epp
\end{multline}

It is now easy to establish a relation with the main result of
\cite{KMST04c}. Define
\begin{equation} \label{transkmst}
     \la_k = \om_k + \e_k \h \epp
\end{equation}
Then
\begin{align}
     & \mspace{-18.mu}
     \frac{1}{n! (m - n)!} \sum_{P \in {\mathfrak S}^m} \sign(P)
        \prod_{1 \le j < k \le m}
	\frac{\sh(\om_{Pk} - \x_j + 2 \e_{Pk} \h) \sh(\om_{Pj} - \x_k)}
	     {\sh \bigl( \om_{Pk} - \om_{Pj} + 2 \e_{Pk} \h \bigr)}
	     \notag \\ & =
     \frac{1}{n! (m - n)!} \sum_{P \in {\mathfrak S}^m} \sign(P)
        \prod_{1 \le j < k \le m}
	\frac{\sh(\la_{Pk} - \x_j + \e_{Pk} \h)
	      \sh(\la_{Pj} - \x_k - \e_{Pj} \h)}
	     {\sh \bigl( \la_{Pk} - \la_{Pj}
	                + (\e_{Pk}  + \e_{Pj})\h \bigr)} \notag \\ & =
     G_n (m|\{\la\}|\{\x\}) \epp
\end{align}
The latter equation defines $G_n (m|\{\la\}|\{\x\})$ which agrees with
the function defined in equation (3.8) of \cite{KMST04c} upon the
identification $\h = \i \z$. Using the formulae in theorem 3.1 and
above lemma 3.2 of \cite{KMST04c} and inserting back (\ref{transkmst})
we arrive at the desired result (\ref{gencoeffint}).

}


\providecommand{\bysame}{\leavevmode\hbox to3em{\hrulefill}\thinspace}
\providecommand{\MR}{\relax\ifhmode\unskip\space\fi MR }
\providecommand{\MRhref}[2]{%
  \href{http://www.ams.org/mathscinet-getitem?mr=#1}{#2}
}
\providecommand{\href}[2]{#2}

\end{document}